\documentclass[DIV=11,a4paper,10pt]{scrartcl}
\usepackage[utf8x]{inputenc}
\usepackage{amsmath}
\usepackage{amssymb}
\usepackage{graphicx}
\usepackage{pictexwd}
\usepackage{color}

\newcommand{\cS}{\mathcal{S}}
\newcommand{\cH}{\mathcal{H}}

\newcommand{\tp}{P}
\newcommand{\tq}{Q}
\newcommand{\cI}{\mathcal{I}}
\newcommand{\cJ}{\mathcal{J}}
\newcommand{\Man}{\mathcal{M}}
\newcommand{\R}{\mathbb{R}}

\newcommand{\C}{\mathbb{C}}
\newcommand{\q}{\mathbf{X}}
\newcommand{\xd}{\mathrm{d}}
\newcommand{\xD}{\mathcal{D}} 

\usepackage{hyperref}

\title{The Unruh-deWitt Detector and the Vacuum in the General Boundary formalism\footnote{AEI-2012-040}}
\author{Ralf Banisch\thanks{ralf.banisch@fu-berlin.de} \\ Freie Universit\"at Berlin \and Frank Hellmann\thanks{frank.hellmann@aei.mpg.de} \\ Albert Einstein Institute \and Dennis R\"atzel\thanks{dennis.raetzel@aei.mpg.de} \\ Albert Einstein Institute}

\begin{document}

\date{\today}

\maketitle

\begin{abstract}
We discuss how to formulate a condition for choosing the vacuum state of a quantum scalar field on a timelike hyperplane in the general boundary formulation (GBF) using the coupling to an Unruh-DeWitt detector. We explicitly study the response of an Unruh-DeWitt detector for evanescent modes which occur naturally in quantum field theory in the presence of the equivalent of a dielectric boundary. We find that the physically correct vacuum state has to depend on the physical situation outside of the boundaries of the spacetime region considered. Thus it cannot be determined by general principles pertaining only to a subset of spacetime.\\

\end{abstract}

\section{Introduction}

The general boundary formulation (GBF) of quantum field theory (QFT), developed extensively in \cite{Oeckl:2003vu,Oeckl2005,Oeckl2006,Oeckl2007,Oe:2dqy,Oeckl2010,Oeckl2011b,Oeckl2012,Oeckl2012b,Oeckl2011a,Oeckl2011,Colosi2011a,CoDo:gen,Co:vac,Co:GBF-QG,Colosi2011,Colosi2009,Colosi2008,Colosi2008a,Colosi2012},
is an axiomatic framework for QFT that allows one to formulate quantum field theory on general spacetime regions with general boundaries. It was originally inspired by, and can be seen as a geometric version of, topological quantum field theory \cite{Atiyah1989a}. More specifically, the set of axioms assigns algebraic structures to geometrical structures and ensures the consistency of these assignments. 

Like Atiyahs axiomatization of TQFTs, which captures the consistency conditions required for a background independent quantum field theory to be sensible, GBF captures the consistency conditions for a quantum field theory on a background metric, but without presuming a 3+1 split, or asymptotic states.

The central subject of the axioms are state spaces associated to boundaries of space time regions and amplitudes for the processes inside them. The question of how states evolve from a Cauchy surface is replaced by the question of whether a state on an arbitrary boundary is extensible (not necessarily uniquely) throughout a space time region.

The GBF amplitudes are a generalization of the standard notion of transition amplitude. As in the standard formulation, probabilities for physical processes can be derived from this amplitude, generalizing the Born rule \cite{Oeckl2007}.

Using the GBF in principle allows the investigation of quantum field theories on compact spacetime regions as well as on regions with purely timelike boundaries. The latter are the subject of this article. We would like to emphasize that neither compact spacetime regions nor those with timelike boundaries can be treated in the standard formulation of QFT.

State spaces and amplitudes for the GBF can be constructed rigorously in different ways, in analogy to different quantization methods developed for ordinary QFT, leading to different representations of the quantum theory. The first construction given in the literature follows a path integral picture and leads to the so-called Schr\"odinger-Feynman representation. It was proposed in \cite{Oeckl:2003vu,Oeckl2005,Oeckl2006}. Another quantization based on geometric and holomorphic quantization for scalar fields leads to the holomorphic representation. This is the representation we will use throughout this paper. It was established for linear field theories in \cite{Oeckl2010} and affine field theories in \cite{Oeckl2011a}. In \cite{Oeckl2011b}, Oeckl showed that there is a one-to-one relation between the two representations, and thus the decision to use one or the other depends only on their respective technical advantages for particular applications.

Furthermore, three different quantization prescriptions for observables in the GBF have been explored so far: one is the Feynman quantization prescription which is a mathematical rigorous version of the path integral quantization of observables \cite{Oeckl2012}, the second is the Berezin-Toeplitz quantization prescription and the third is the normal ordering prescription \cite{Oeckl2011}. Among them, the Feynman quantization prescription is preferred as it is used in standard quantum field theory, leading to results that agree with experimental observations. Furthermore, in contrast to the other quantization prescriptions, the Feynman quantization prescription fulfills natural identities when several spacetime regions are combined \cite{Oeckl2012}.

In this article we study in detail physical conditions one can impose to determine the remaining ambiguities in the quantization, that is, the choice of vacuum state, or in the language of holomorphic quantization, the complex structure. This condition is based on coupling an Unruh-DeWitt detector moving along a specified world line in Minkowski space and studying the transition rates given certain boundary states.

We find that, if a notion of outgoing and ingoing modes is specified in advance, the condition that a detector at rest does not click for outgoing modes can fix the complex structure for the ordinary field modes, i.e. propagating waves.

For field modes that behave exponentially in space, i.e. evanescent waves, it is less clear how to select the states for the no-click condition. Thus we also study the actual transition rates in a concrete experimental set up capable of producing modes that are exponentially decaying in space \cite{carniglia1971,carniglia1972}.

This is done by putting a dielectric boundary outside of the timelike region we consider. We find that we can reproduce the result for the response of the detector derived from canonical quantization in the GBF only if we take the physics outside the spacetime region into account. Then, the corresponding complex structure agrees with the one obtained from asymptotic consideration. This can be seen as a GBF manifestation of the non-locality of the vacuum state.

\subsection{Outline of the article}

To introduce the reader to the general boundary formulation of quantum theory we consider the quantum harmonic oscillator formulated in the language of the GBF in the following section. Then we consider holomorphic boundary states and amplitudes. We call this the boundary formulation of the quantum harmonic oscillator skipping the word ``general`` since in the corresponding set up there is only one possible choice of boundary. 

We then briefly review the case of a scalar general boundary quantum field theory in Section \ref{sec-Quant-ComplexStruct} in the holomorphic quantization scheme, discussing in particular the notion of unitarity in the theory, and the construction of one-particle-states and the one point function. The latter allows us to give the GBF formulation of the Unruh-DeWitt detector by using the text book formulas for the detector coupling. An independent derivation that happens fully in the GBF framework is given in appendix \ref{sec:detector}, using coherent states which are reviewed in appendix  \ref{sec-CoStat}.

In Section \ref{sec:noclick}, we show that, using symmetry arguments and an appropriate choice of boundary states, we can determine the complex structure for propagating modes by requiring a detector at rest to not click. In Section \ref{sec:ExperimSituation}, we then consider an actual physical example in which evanescent waves appear as the appropriate choice of boundary state is unclear for evanescent waves using the general boundary formulation. This example is a toy model of a free electromagnetic field in the presence of a dielectric boundary. The excitation rates induced by evanescent waves in this case have been studied theoretically and experimentally in \cite{carniglia1971,carniglia1972}. We show that in order to reproduce the transition rates while maintaining the ability to model the two situations of particle emission and particle absorption as orthogonal subsets in the boundary Hilbert space, which is the situation one has in canonical quantization, the complex structure has to depend on the physical situation at the dielectric boundary, even if this boundary is outside of the spacetime region considered. We conclude that a complex structure for evanescent waves cannot be chosen by general physical arguments pertaining only to the spacetime region under consideration. We further show that the correct result for the transition rate of the detector is reproduced if a particular non-unitary complex structure is considered that reproduces the right asymptotic behavior. Such a non-unitary complex structure is used for the first time in interacting GBQFT in this article.

\section{The quantum harmonic oscillator} \label{sec:harmosc}

In order to introduce the ideas underlying the general boundary formalism as well as those of holomorphic quantization we will begin by discussing these ideas using the harmonic oscillator as a concrete example for a system with one degree of freedom. Here we interpret the harmonic oscillator as a 0-dimensional quantum field theory, that is, a quantum system on a world line. The boundaries of a segment of a world line are simply its start point, at parameter $t_i$ and end point at parameter $t_f$.  In the following section we will then introduce the more general constructions used for the scalar field. The harmonic oscillator itself will also reappear as an Unruh-deWitt type detector in Section \ref{sec:detector}.

\subsection{The GBF formulation}

While the standard formulation of quantum mechanics is concerned with the evolution of a system in the Hilbert space as it changes through time $t$, the general boundary formulation of quantum mechanics considers amplitudes between initial and final states. For 0-dimensional systems this is not more general, merely a rewriting that suggests an obvious generalization for the higher dimensional case.

To make this concrete for the 0-dimensional case we introduce a copy of the Hilbert space of the harmonic oscillator at each time $t$. These spaces are all canonically identified with the space at some fiducial time $t=0$, $\cH_t := \cH_0$. We have the standard unitary evolution from $\cH_t$ to $\cH_{t'}$ given by $U(t',t)$. The usual statements we want to make in the theory concern probabilities to observe the system in a subspace $\cS_f \subset \cH_{f} := \cH_{t_f}$ given that it was prepared in a subspace  $\cS_i \subset \cH_{i} := \cH_{t_i}$. We will denote the projectors on the subspace $\cS_{i/f}$ by $P_{i/f}$. We introduce an orthonormal basis of $\cS_{i/f}$ as $\psi^a_{i}$ and $\psi^a_{f}$. We model this situation in the text book way by taking the system at time $t_i$ to be given by the density matrix $\frac{P_i}{tr P_i}$, then at time $t_f$ it will be in the state $$\sigma = U(t_f, t_i) P_i U(t_i, t_f)/tr(P_i).$$ The text book probability to observe the proposition $P_f$ is then given by $tr(\sigma P_f)$ or more explicitly

\begin{equation}\label{eq:StandardProbability}
P(\cS_f|\cS_i) = \frac{tr(P_i U(t_i, t_f) P_f U(t_f, t_i))}{tr(P_i)} = \frac{\sum_{a,b} |{\psi_f^b}^\dagger U(t_f,t_i) \psi_i^a|^2}{\sum_a |\psi_i^a|^2}.
\end{equation}

The general boundary formalism takes this as its starting point. We take the boundary Hilbert space $\cH_{\partial I}$ of a segment $I = [t_i,t_f]$ to be the space $\cH_i \otimes \cH_f^*$, where $*$ indicates the Hermitian dual. Then $\cS_i \otimes \cS_f^* = \cS_\cJ$ is the subspace of events we are interested in, relative to the preparation subspace $\cS_i \otimes \cH_f^* = \cS_\cI$, which only encodes the information on how the system was prepared. Note that here we have $\cS_\cJ \subset \cS_\cI$.

We want to express the above formula given the spaces $\cS_{\cI}$ and $\cS_{\cJ}$ directly. To do so it is convenient to consider the unitary evolution $U$ as a map from the boundary space to $\C$, that is, we introduce $\rho_I: \cH_i \otimes \cH_f^\star \rightarrow \C$ as the linear map defined by

\begin{equation}
\rho_I(\psi_i \otimes \psi_f^\dagger ) = \psi_f^\dagger U(t_f,t_i) \psi_i.
\end{equation}

$\rho_I$ is simply $U(t_f,t_i)$ seen as an element of $ \cH_i^\star\otimes\cH_f = \cH_{\partial I}^*$. The map $\rho_I$ can then also be read as the inner product in $\cH_{\partial I}$, of the normalized state $\psi = \psi_i \otimes \psi_f^\dagger$ with the state $U^\dagger = \rho_I^\dagger \in \cH_{\partial I}$. That is,

\begin{equation}
\rho_I(\psi) = \langle \rho_I^\dagger | \psi \rangle_{\partial I} = \langle U^\dagger | \psi \rangle_{\partial I}.
\end{equation}

Viewed this way the norm square of the map $\rho_I$ can be seen as the length of the projection of $\rho_I^\dagger$ on the subspace spanned by $\psi$, that is, writing the projector onto that subspace $P_\psi$ we have that $$|\rho_I(\psi)|^2 = |P_\psi \rho_I^\dagger|^2.$$

This allows for an immediate generalization to arbitrary subspaces $\cS$. Given a projector $P$ onto $\cS$ and an arbitrary orthonormal basis $\psi_c$ of $\cS$, the amplitude can thus be written as $$|\rho_I \circ P|^2 = |P \rho_I^\dagger|^2 = \sum_c |\rho_I(\psi_c)|^2.$$

It is then straightforward to see that Equation \eqref{eq:StandardProbability} can be written as

\begin{equation}\label{eq:GBFProbability}
P_{\text{standard}}(\cS_f|\cS_i) = P_{\text{GBF}}(\cS_{\cJ}|\cS_{\cI}) = \frac{|\rho_I \circ P_{\cJ}|^2}{|\rho_I \circ P_\cI|^2}.
\end{equation}

The central assumption of the general boundary formalism is that the above formula remains valid for wide classes of boundaries, subspaces and amplitude maps $\rho_I$ in higher dimensions.

\subsubsection{Example: The harmonic oscillator}\label{sec:HarmOsci1}

For the harmonic oscillator with classical variables $q$, $p$ we have the usual Hilbert space $\cH_0 = L^2(\R,dq)$. We write $\cH_t = L^2(\R,dq_t)$. The usual operators are 
\begin{equation}
 \hat p \psi(q)=-i\frac{d}{dq}\psi(q), \quad \hat q \psi(q)=q\psi(q)
\end{equation}
and, using the shorthand $M = \sqrt{m\Omega}$
\begin{equation}
 \hat a:= \frac{1}{\sqrt{2}}\left( M \hat q+i \frac{1}{M}\hat p\right),\quad \hat a^\dagger:=\frac{1}{\sqrt{2}}\left(M \hat q - i \frac{1}{M}\hat p\right)\,,
\end{equation}
with $H = \Omega (\hat{a}^\dagger \hat{a} + \frac12)$, and the evolution map $U(t_f - t_i) = \exp(i (t_f - t_i) H)$.

From this we immediately have the general boundary construction above with $$\rho_{I}(\psi_i \otimes \psi_f^\dagger) = \psi_f^\dagger \exp(i (t_f - t_i) H)\psi_i.$$

\subsection{Holomorphic quantization}

For us it will be interesting to see an alternative construction of the general boundary formulation ingredients though, given by holomorphic quantization. This can be seen as a special case of geometric quantization \cite{woodhouse1997geometric} and was introduced for affine and linear general boundary field theory in \cite{Oeckl2010,Oeckl2011a}.

In the usual description of the harmonic oscillator above, we represent the algebra of observables by acting on the space of functions on just one of the canonical variables, $q$. These can be seen as functions on phase space satisfying $\frac{d \psi}{dp} = 0$. We can instead work with a different set of functions on phase space that instead satisfy $(M^{-1}\frac{d}{dq} - i M\frac{d}{dp}) \psi = 0$. Introducing $\tq = M q$ and $\tp = M^{-1} p$, and the variable $a=\tq + i \tp = M q+i M^{-1} p$, these are exactly the holomorphic functions of $a$. These are the functions that satisfy $\frac{d\psi}{d\overline{a}} = 0$, where $\frac{d}{d\overline{a}}$ is the Wirtinger derivative $\frac{d}{d\overline{a}} = \frac{d}{d\tq} - i \frac{d}{d\tp}$.

This space of holomorphic functions on the complex line, $$L^2_{\text{hol}}(\C,d_{\text{Gauss}}),$$ with inner product
\begin{equation}
\langle \psi,\psi'\rangle:=\int_{\C}\frac{da}{2\pi}\,e^{-\frac12|a|^2} \overline{\psi(a)}\psi'(a),
\end{equation}
is particularly well suited for the harmonic oscillator and the bosonic scalar field.

The representation of the algebra of observables is easily given in terms of creation and annihilation operators $\hat{a}^\dagger$ and $\hat{a}$. On vectors $\psi(a) \in L^2_{\text{hol}}(\C)$ these are defined simply as multiplication and derivative operators respectively,

\begin{equation}
\hat{a}^\dagger \psi(a) = \frac{a}{\sqrt{2}} \psi(a), \quad \hat{a} \psi(a) = \sqrt{2} \frac{d}{da}\psi(a),
\end{equation}
where $\frac{d}{da}$ is again a Wirtinger derivative $\frac{d}{da} = \frac{d}{d\tq} + i \frac{d}{d\tp}$.

It is immediate that $\hat{a}$ and $\hat{a}^\dagger$ satisfy the commutation relations of the creation and annihilation operators. To check whether this is indeed a representation, we need to check that the hermitian dual of $\hat{a}$ with respect to the Gaussian inner product above is indeed $\hat{a}^\dagger$, which follows from partial integration, and the fact that the Wirtinger derivative of the antiholomorphic function $\overline{\psi}$ vanishes.

We can now also see how the holomorphic picture is well suited to the harmonic oscillator. The constant function $\psi_0(a) = 1$ is annihilated by $\hat{a}$ and thus gives the ground state. General states generated by $\hat{a}^\dagger$ and $\hat{a}$ are simply polynomials of $a$. Coherent states take the simple form, $$K_z(a) := \exp(\frac1{\sqrt{2}} z \hat{a}^\dagger) \psi_0(a) =  \exp(\frac12\overline{z} a) .$$
They are of particular interest because they fulfill the reproducing kernel property
\begin{equation}\label{eq:harmoscrepkern}
 \langle K_z,\psi\rangle=\psi(z)\quad\text{for any}\,\psi(a)\in L^2_{hol}(\C)\,.
\end{equation}
This can be seen by expressing $\psi(a)$ in a Taylor series, expressing the powers of $a$ with powers of the creation operator $\hat a^\dagger$ and the vacuum state, using the duality of $\hat a$ and $\hat a^\dagger$, shifting the integration variable $a$ to $a - z$ and using that $\langle \psi_0,\psi_0 \rangle=1$.

\subsubsection{Complex structures}

The space of holomorphic functions on phase space that is used depends on the choice of complex structure on phase space which provides an identification of phase space with the complex numbers. Different complex structures will lead to different function spaces.

This ambiguity is of no big consequence in finite systems. The Stone-von Neumann theorem ensures that all representations of the canonical commutation relations are unitarily equivalent, even though this might not be obvious in practice. This will change in the case of the scalar field that we will consider later in this paper, where the choice of complex structure can be shown to correspond to a choice of vacuum state. For this reason we will here discuss several different ways in which this choice of complex structure can be encoded.

The classical data which we have to start with is $L$, the phase space of our system, parametrized by canonically conjugate coordinates $p$,$q$, and equipped with an antisymmetric bilinear form $\omega$, the Poisson bracket. We call the complexification of this phase space, $L^\C$.

The identification with the complex numbers can be encoded by specifying how the multiplication with the imaginary unit acts on phase space. That is, if $\pi$ is the real-linear map from $\C$ to $L$, we are interested in the operator $J$ defined by

$$J\circ \pi = \pi \circ i.$$

\begin{figure}[htb]
 \begin{center}
  \includegraphics[scale=0.2]{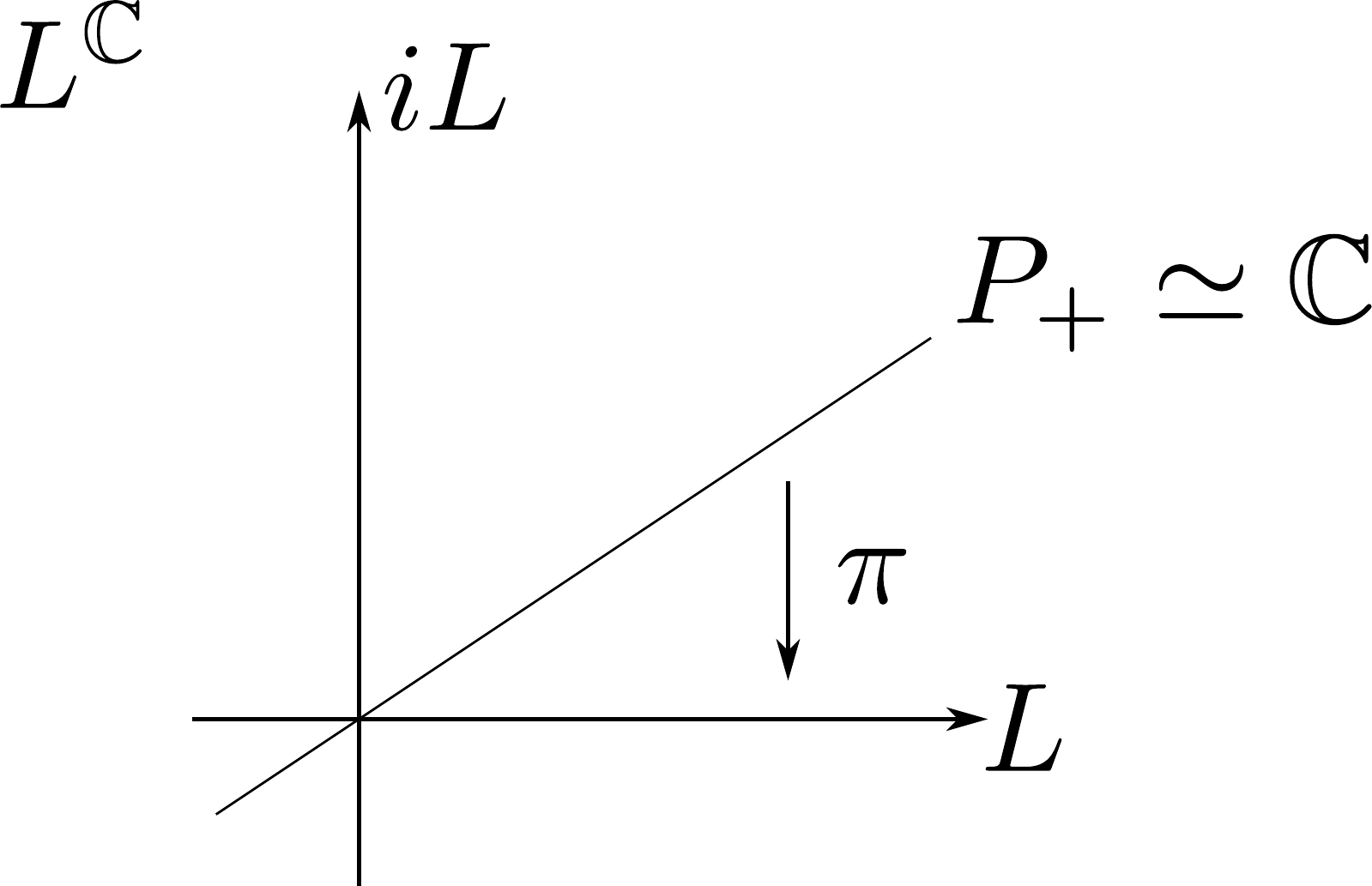}
  \caption{The projection $\pi$ from $P_+$ to the real subspace $L \subset L^\C$.}
 \label{fig:Projection}
 \end{center}
\end{figure}

This operator is called a complex structure and has the property the $J^2 = -1$. Thus it has eigenvalues $\pm i$. We can diagonalize it by lifting it to the complexified phase space $L^\C$. Its eigenspaces $P_\pm$ are complex linear subspaces of $L^\C$ of complex dimension one, as $P_+ = \overline{P_-}$. The projection on its $+i$ eigenspace $P_+$, $$\pi^{-1} = \frac12 (1 - i J),$$ provides us with an immediate identification of the real subspace $L$ and $P_+ \simeq \C$. Its inverse is always just given by $\pi(\cdot) = \Re(\cdot)$. Thus a complex structure does indeed provide us with an identification of phase space with $\C$, as claimed. This is illustrated in Figure \ref{fig:Projection}.

If the complex structure satisfies $\omega(J \cdot, J \cdot) = \omega(\cdot, \cdot)$, and the symmetric linear form $g(\cdot,\cdot) = \omega(J \cdot, \cdot)$ is positive, we call $J$ compatible with $\omega$. The triple $(\omega, J, g)$ is called a K\"ahler structure on phase space. We will always require that $\omega$ and $J$ define such a K\"ahler structure. This ensures that the inner product on $P_+ \subset L^\C$, defined in terms of $g$ and $\omega$ as $\{ \cdot,\cdot\} = g(\pi(\cdot),\pi(\cdot)) + 2 i \omega(\pi(\cdot),\pi(\cdot))$, is Hermitian.

A more thorough and precise introduction to these structures in the general case will be given in Section \ref{sec-Quant-ComplexStruct}.

\subsubsection{Harmonic oscillator continued...}\label{sec:HarmOsc2}

We will now continue looking at the example of the holomorphic harmonic oscillator from Section \ref{sec:HarmOsci1} and write down the above structures in this case.

We have $a = M q + \frac{i}{M}p$ so, introducing the shorthand notation $\q = (p,q)^T$, the map $\pi$ from $L$ to $\C$ is explicitly given by $\pi^{-1}(\q) = Mq + \frac{i}{M}p$. Then $J$ is the matrix

$$J = \left(\begin{matrix} 0 & -1/M^2 \\ M^2 & 0 \end{matrix}\right).$$

The standard symplectic structure inherited by the dynamics is in these coordinates 

$$\omega(\q',\q) = \frac{1}{2}(pq' - qp') = (q',p') \left(\begin{matrix} 0 & 1/2 \\ -1/2 & 0\end{matrix}\right)\left(\begin{matrix} q \\ p \end{matrix}\right).$$

It is straightforward to see that $\omega(J\cdot,J\cdot) = \omega(\cdot,\cdot)$, and that the bilinear form 
\begin{equation}
g(\q',\q) := 2\omega(\q',J\q) = (q',p')^T \left(\begin{matrix} M^2 & 0 \\ 0 & 1/M^2 \end{matrix}\right) \binom{q}{p}
\label{eq_metric}
\end{equation}
is positive and symmetric and thus defines a natural metric on phase space. Thus $\omega, J$ and $g$ form a K\"ahler structure as required. As noted above they define a complex inner product $\{ \cdot,\cdot\}$ via

$$\{ \q',\q\} := g(\q',\q) + 2i\omega(\q',\q).$$

This complex inner product is sesquilinear in the sense that $\{ \q',J\q\} = i\{ \q',\q\} = -\{ J\q',\q\}$. In this case it is 

$$\{ \q',\q\} = \left(Mq'-\frac{i}{M}p'\right)\left(Mq+\frac{i}{M}p\right) = \bar{a}'a,$$

the familiar sesquilinear form on $\C$.
The eigenspaces of $J$ in $L^\C \simeq \C^2$ are given by

\begin{equation*}
P_\pm = \left\{ \left(\begin{matrix} z \\ \mp iM^2 z \end{matrix}\right), z \in \C\right\}.
\end{equation*}
More generally, the most general linear map on $L$ which satisfies $J^2=-1$ can be written as
\begin{equation}
J = \left(\begin{matrix} b & -c^{-1}(1+b^2)^{\frac12} \\ c (1+b^2)^{-\frac12} & -b\end{matrix}\right).
\label{J_gen}
\end{equation}
It can be shown that $J$ and $\omega$ are compatible, or equivalently that $g(\cdot,\cdot) = 2\omega(\cdot, J\cdot)$ is symmetric, and that $g$ is positive, iff we have $(b,c) \in \R \times (0,\infty)$. The choice we made at the beginning of this section corresponds to $b=0$ and $c = M^2$.

The eigenspace $P_+$ for the complex structure (\ref{J_gen}) can be written as
\begin{equation*}
 P_+ = \left\{ \left(\begin{matrix} z \\ e^{-i\phi} c z \end{matrix}\right), z \in \C\right\}
\end{equation*}
where $e^{-i\phi} = (b-i)(ab^2+1)^{-1/2}$, in other words $\phi\in [0,\pi]$.

\subsection{Comparison of complex structures}\label{sec:compcompl}

Different complex structures lead to different identifications of phase space with the complex numbers. Thus the spaces of holomorphic functions for different complex structures, when viewed as functions on phase space, differ. However, it is still possible to compare functions in theses different function spaces, by considering them as square integrable functions on phase space and taking an inner product there. This inner product is called a pairing \cite{Woodhouse:1981zd,woodhouse1997geometric}.

Concretely, given two different identifications, $a(p,q) = \pi^{-1}(p,q) \neq a'(p,q) = \pi'^{-1}(p,q)$, and states $\psi(a)$, $\psi'(a)$ in $L^2_{\text{hol}}(\C,d_{\text{Gauss}}a)$ and $L^2_{\text{hol}}(\C,d_{\text{Gauss}}a')$ respectively, we can construct functions $\phi(p,q) = e^{-\frac14|a(p,q)|^2} \psi(a(p,q))$, and $\phi'(p,q) = e^{-\frac14|a'(p,q)|^2} \psi'(a'(p,q))$ that are square integrable on phase space. That is, $\phi, \phi' \in L^2(L, dqdp)$.\footnote{In general the volume element $dqdp$ is given by the symplectic form.}

We denote the pairing between $\psi$ and $\psi'$ by $(\psi, \psi')_p$. Explicitly it is given by

\begin{equation}
(\psi,\psi')_p = \int dq dp e^{-\frac14|a(q,p)|^2-\frac14|a'(q,p)|^2} \overline{\psi(a(q,p))} \psi'(a'(q,p)).
\end{equation}

This pairing has different physical interpretations that need to be established on a case by case basis.

\subsubsection{Pairings for the harmonic oscillator}

Note that as any classical time evolution $T(t,t')$ generated by a Hamiltonian vector field leaves the symplectic form $\omega$ invariant, the operator $J' = T(t,t') \circ J \circ T(t,t')^{-1}$ is also a compatible complex structure. In this case the classical time evolution operator $T(t,t')$, considered as a map between the complex spaces arising from $J$ and $J'$, is unitary. As the hermitian inner product on these complex spaces is defined in terms of $\omega$ and $J$, this will always be the case if we have $J'\circ T(t,t') = T(t,t') \circ J$.

Remarkably, for the harmonic oscillator the pairing between the holomorphic Hilbert spaces at different times provides us with the full quantum dynamics. To see this we can view phase space as the space of solutions $x(t)$ to the equations of motion,

\begin{equation}\label{eq:harmparam}
{x}(t)= \frac{1}{2 M}\left(a_0 e^{-i\Omega t}+\overline{a_0} e^{i\Omega t}\right)\,.
\end{equation}

with $$a_0 = M q + i\frac{p}{M}\,,$$ given in terms of our previous phase space coordinates. We then have an entire family of phase space coordinates, $$q_t = x(t) \,\;\;\text{and}\,\;\; p_t = m \frac{d x}{d t} (t),$$ and $$a_t = M q_t + i\frac{p_t}{M}\, = \pi^{-1}_t(q,p),$$ provides a whole family of complex structures on phase space. Note that the operators $J$ can be written in very compact form as

\begin{equation}
J = \frac{\partial_t}{\Omega},
\end{equation}

Expressing this operator in the various coordinates $q_t$ and $p_t$ provides the complex structures $a_t$.

These complex structures are related quite simply by $a_t = e^{i \Omega (t - t')} a_{t'}$. As anticipated it is a unitary transformation. The pairing is now of the form

\begin{equation}
(\psi_t,\psi_{t'})_{\text{pairing}} = \int dqdp e^{-\frac14 |a_t|^2 - \frac14 |a_{t'}|^2} \overline{\psi_t(a_t)} \psi_{t'}(a_{t'}),
\end{equation}

which, noting that $|a_t|^2 = |a_{t'}|^2$ are real and thus equal to $g^2 = g((q,p),(q,p))$, we can simplify to 

\begin{equation}
(\psi_t,\psi_{t'})_{\text{pairing}} = \int dqdp e^{-\frac12 g^2} \overline{\psi_t(e^{i \Omega t}a_0)} \psi_{t'}(e^{i \Omega t'}a_{0}).
\end{equation}

We can now check that we actually have

\begin{equation}\label{eq:harmoscunitary}
(\psi,\psi')_{\text{pairing}} = \langle \psi | U(t - t') \psi' \rangle,
\end{equation}

where $\langle| \rangle$ is the inner product in $L^2_{\text{hol}}(\C,d_{\text{Gauss}})$, with $U(t - t') = \exp(-i \Omega  \bar a a (t - t'))$.

Thus, remarkably, the pairing between the Hilbert spaces associated to the different complex structures is related by the classical time translation and provides us with the full quantum dynamics. In general the pairing can be identified with semi-classical transition amplitudes. It is only for the harmonic oscillator and linear field theory, where it provides us with the Bogoliubov transformation, that the pairing provides the full quantum mechanical amplitudes.

The pairing can be evaluated very simply on coherent states, using the reproducing kernel property in Equation (\ref{eq:harmoscrepkern}) and the unitarity relation in Equation (\ref{eq:harmoscunitary}) to obtain
\begin{equation}
(K_{z,t},K_{z',t'})_{\textrm{pairing}} = e^{\frac{1}{2} \overline{z'} z}=K_{z'}(z)\,.
\end{equation}

This ability to solve the dynamics in a straightforward way which makes the dependence on the underlying choices of complex structure obvious will also be true for the case of the field theory. This is what motivates our use of holomorphic quantization.

Note that the operator ordering in the quantum Hamiltonian is not the standard one for the harmonic oscillator. It differs by the ground state energy of $\frac12 \Omega$. This is the standard ordering for the free field theory, and the discrepancy for the harmonic oscillator will not matter for our application. The correct ordering can be obtained at the cost of significant technical complications, and we refer the interested reader to the literature, in particular the book \cite{woodhouse1997geometric}, on this point.

\subsection{The holomorphic general boundary harmonic oscillator}

With the help of the pairing we can now give a very simple prescription for constructing the holomorphic general boundary harmonic oscillator for an interval $I = [t_i,t_f]$ from the classical data.

The space of solutions for the harmonic oscillator is parametrized by the position and momentum at a certain time $t$, which, without adding any extra structure, we call $L_{t} \sim \R^2$. Given two copies of this phase space, one given by the parametrization $L_{i}$ at time $t_i$ and one by the parametrization $L_{f}$ at time $t_f$ we can form the direct sum of these spaces $L_i \oplus L_f$ to obtain the boundary phase space. The physical phase space $L_{p}$ is then simply a 2-dimensional subspace of this boundary phase space with elements $(q_i,p_i,q_f,p_f)$ such that $(q_i,p_i) \in L_i$ parametrizes the same solution as $(q_f,p_f) \in L_f$.

We can now directly quantize this setup using holomorphic quantization to obtain the general boundary formulation of the harmonic oscillator in this language. We put a complex structure on $L_i$, take the time translation related one at $L_f$, identify them with $\C$ and use the holomorphic Hilbert space construction to define the boundary Hilbert space for the region $I = [t_i,t_f]$, that is, $\cH_{\partial I} = \cH_i \otimes \cH_f^*$ with $\cH_{i/f} = L_{\text{hol}}$. We then use the pairing to define an amplitude map $\rho_I$

\begin{equation}
\rho_I (\psi' \otimes \psi^\dagger) := (\psi,\psi')_{\textrm{pairing}} = \int dq dp e^{-\frac12g^2} \overline{\psi(a_t)} \psi'(a_{t'})\,,
\end{equation}
where $g^2 = g((q,p),(q,p))$.

\subsection{The higher dimensional case}

The structure we considered above was for the $0$-dimensional quantum field theory. The prescription of general boundary field theory for higher dimensions is to consider as classical boundary spaces, the space of solutions in the neighborhood of the boundary. The question of dynamics is then coded in the question if a particular solution near the boundary can be extended to the bulk.

This immediately generalizes our discussion above, as our boundary consists of two parts, which each individually corresponds to solutions near the boundary that are extensible throughout time. Thus the question of dynamics is simply if the extension from both parts of the boundary defines the same solution. This is another way to describe the subspace $L_p \subset L_f \oplus L_i$ given above.

In what follows we will always have a boundary with two components. We will always take the space of classical solutions near each of these components to be the space of solutions near them that can be extended throughout space time, that is, phase space. Thus the space of solutions near the entire boundary is the direct sum of two copies of phase space.

To then move on to construct a quantum theory based on this classical data we will need to again specify complex structures and construct Hilbert spaces on the spaces $L$. We will now do so for the special case of the Klein-Gordon field.

\section{The holomorphic GBF Klein-Gordon field and the Unruh-DeWitt detector}\label{sec-Quant-ComplexStruct}

In this section we give the analogues structures to the holomorphic boundary formulation of the harmonic oscillator for the Klein-Gordon field, and give the Unruh-DeWitt detector coupling in this formulation. The key new element is the ambiguity in the choice of K\"ahler manifolds and K\"ahler polarizations. The holomorphic quantization scheme for this case was given in detail in \cite{Oeckl2010,Oeckl2011a} for linear and affine field theories. It is a special case of geometric quantization that can be found in detail in \cite{woodhouse1997geometric}. In section \ref{sec:ladder} we give the one particle sector and the one point function. In \ref{sec:timeev} we will define the general notion of unitarity we already found for the harmonic oscillator above. In Section \ref{sec:symmred}, we will show how to impose spacetime symmetries on the complex structure, this is what usually singles out a preferred complex structure/vacuum state in Minkowski space, and leads to reduction of the problem of defining the complex structure to energy eigenspaces in our case. Finally in section \ref{sec:UdW} we give the Unruh-DeWitt detector response rates.

Holomorphic general boundary quantization in general spacetime regions with general boundaries is a difficult, and unsolved mathematical problem. In order to make it tractable for us, we will always make the following assumptions throughout the rest of the paper:

\begin{enumerate}
 \item[(A1)] We consider piecewise flat spacetimes $\mathcal{M}$ which are differentiable manifolds of dimension 2 and 4 equipped with a Lorentzian metric $g$ with signature $(+-)$ and $(+---)$ respectively.
 \item[(A2)] Regions $M\subseteq \mathcal{M}$ are assumed to be orientable submanifolds with a two-component and flat boundary $\Sigma = \Sigma_1 \cup \Sigma_2$ with orientation inherited from $M$ and such that $\Sigma_1 \cap \Sigma_2 = \emptyset$. We choose the unit normal vector field $n$ on $\Sigma$ to be outward pointing. $\Sigma$ is either everywhere spacelike or everywhere timelike. Thus in particular $\Sigma_1$ and $\Sigma_2$ are parallel.
\item[(A3)] We only consider solutions to the field equations which have the following global extension property: If a solution is specified in a neighbourhood of $\Sigma_{1/2}$, it can be uniquely extended to all of $M$. 
\end{enumerate}

In particular, in section 4 we will consider a two-dimensional Minkowski strip for explicit calculations.

\subsection{Holomorphic GBF of the Klein Gordon field}

To give the Klein-Gordon fuel in the holomorphic GBF we need to classical phase space structure first. For any region $M\subseteq \Man$ we have the action integral

$$S_M(\phi) := \int_M \sqrt{-g} L(\phi(x),\partial\phi(x)) dx$$

where $L(\phi,\partial\phi)$ is the usual Lagrangian density for the Klein-Gordon field on a Lorentzian spacetime
\begin{equation}
 \label{eq:action}
 L(\phi,\partial\phi,x):=\frac{1}{2}g^{\mu\nu}\partial_\mu\phi(x)\partial_\nu\phi(x)-\frac12 m^{2}\phi^2(x)\,,
\end{equation}


As noted before, the GBF uses the real vector space $L_{\Sigma}$ of solutions to the Euler-Lagrange-equations in a neighborhood of $\Sigma$. The dynamics can than be phrased as the problem of extending the solutions from a neighborhood of $\Sigma$ to $M$.

The symplectic form can be derived from the action and is given by

\begin{equation}
\omega_{\Sigma}: L_{\Sigma}\times L_{\Sigma}\rightarrow \R, \quad \omega_{\Sigma}(\phi,\phi') = \frac{1}{2}\int_{\Sigma} \sqrt{|\det g(x)|} n^\mu(x) \left(\phi'(x) \frac{\partial L}{\partial(\partial_\mu \phi)}(x) - \phi(x) \frac{\partial L}{\partial(\partial_\mu \phi')}(x) \right)\,,
\end{equation}
where $n$ is the outward unit normal vector field to the hypersurface $\Sigma$.

If $\Sigma = \Sigma_1 \cup \Sigma_2$ as in (A2), we have $L_{\Sigma} = L_{\Sigma_1}\oplus L_{\Sigma_2}$, and $\omega_{\Sigma} = \omega_{\Sigma_1} + \omega_{\Sigma_2}$. Note that for $\overline{\Sigma}$, that is $\Sigma$ with the opposite orientation we have $\omega_{\overline{\Sigma}} = - \omega_{\Sigma}$. For details of this construction, and proofs of the preceding statements, we refer again to the literature \cite{woodhouse1997geometric}.

{In general the symplectic form does not need to be non-degenerate. Per \cite{woodhouse1997geometric} this degeneracy is related to the presence of gauge orbits, or equivalently, the non uniqueness of the extension of solutions from the neighbourhood of a surface to the entire space time. Due to our assumption A3 this is not the case here, and we will assume non-degeneracy from here on.}

By (A3), we obtain from $\omega_{\Sigma_{1/2}}$ a symplectic structure $\omega_{M_{1/2}}$ on the space $L_M$ of global solutions on $M$. Furthermore, we assume that $\omega_{M_1}$ and $\omega_{M_2}$ coincide and just call both $\omega_M$.  
This means that $L_{\Sigma} \simeq L_{M} \oplus L_{M}$, and $$\omega_\Sigma((\phi_1,\phi_2),(\phi'_1,\phi'_2)) = \omega_M(\phi_1,\phi_1') - \omega_M(\phi_2,\phi'_2),$$ where $\phi_{1/2} \in L_M$. This is of course just the case we described for the Harmonic oscillator. 

Note that if we have an element $(\phi_1,\phi_2)\in L_\Sigma$ that extends to a solution on the interior of $M$ we have that $\phi_1$ and $\phi_2$ denote the same element of $L_M$, and thus elements $\phi$, $\phi' \in L_{\Sigma}$ that extend to the interior of $M$ have $\omega_{\Sigma}(\phi,\phi') = 0$. Therefore $L_M$ forms a Lagrangian subspace of $L_{\Sigma}$. This property is actually not specific to our case, but true more generally, we refer the interested reader to the literature \cite{Oeckl2010}.

We now pick a {\em polarization} $\mathcal{P}$ of $L_{\Sigma}^{\C}$, that is a Lagrangian subset (the 'subset of positive frequency modes') $\mathcal{P} \subset L_{\Sigma}^{\C}$ of the complexification of $L_{\Sigma}$. If $\mathcal{P}$ is a polarization, then so is $\overline{\mathcal{P}}$, where the bar denotes complex conjugation in the usual sense. $\mathcal{P}$ is called a {\em K\"ahler polarization} if $\mathcal{P}\cap\overline{\mathcal{P}} = \{0\}$. In this case, we can write $L_{\Sigma}^{\C} = \mathcal{P} \oplus\overline{\mathcal{P}}$.

K\"ahler polarizations are in one-to-one correspondence with complex structures $J_{\Sigma}$ compatible to $\omega_{\Sigma}$ by

\begin{eqnarray*}
 \mathcal{P} & := & \{\phi\in L_{\Sigma}^{\C} | J_{\Sigma}\phi = i\phi\}\\
 \overline{\mathcal{P}} & := & \{\phi\in L_{\Sigma}^{\C} | J_{\Sigma}\phi = -i\phi\}
\end{eqnarray*}

For further details see \cite{1999math.ph...4008E}. We further always assume $J_\Sigma$ to be a positive compatible complex structure.

As in the case of the harmonic oscillator the sesquilinear form

$$\{\phi,\eta\}_{\Sigma} := g_{\Sigma}(\phi,\eta) + 2i\omega(\phi,\eta)\quad\forall\phi,\eta\in L_{\Sigma}$$

turns the real vector space $L_{\Sigma}$ into a complex Hilbert space, where multiplication with $i$ is given by applying $J_{\Sigma}$. Real linear maps will lift to complex linear maps exactly if they commute with $J_{\Sigma}$. We have a canonical isomorphism $\pi: \mathcal{P} \rightarrow L_{\Sigma}$, $\pi(\psi) = \psi + \bar\psi$ that identifies $\mathcal{P}$ and $L_{\Sigma}$, the latter now regarded as a complex vector space. The inverse is explicitly given in terms of the complex structure as

$$\pi^{-1} = \frac12 (1 - i J_\Sigma).$$

In the case of the scalar field $L_{\Sigma}$ as a complex space can be interpreted as the (dual of the) one-particle Hilbert space. This is analogous to the situation of the Harmonic oscillator, there the  complex classical phase space is simply $\C$, and can be identified with the one dimensional space of first excitations. 

The state space $\cH_{\Sigma}^h$ of the holomorphic quantization is now constructed as\footnote{To deal with technical difficulties arising in the infinite dimensional case, one actually has to construct $\cH_{\Sigma}^h := L^2_{hol}(\hat L_{\Sigma},d\mu_{\Sigma})$, where $\hat L_{\Sigma}$ is a certain extension of $L_{\Sigma}$, namely the algebraic dual of its topological dual. The full construction of $\hat L_{\Sigma}$ and $d\mu_{\Sigma}$ is given in \cite{Oeckl2010}.} $\cH_{\Sigma}^h := L^2_{hol}(L_{\Sigma},d\mu_{\Sigma})$ the space of square-integrable holomorphic functions on $L_{\Sigma}$ with respect to the inner product

$$\langle \psi',\psi\rangle_{\Sigma} :=\int_{L_{\Sigma}}\psi(\phi)\overline{\psi'}(\phi) d\nu(\phi)\,,$$

where $d\nu$ is a Gaussian measure determined by $\frac{1}{2}g_{\Sigma}(\cdot,\cdot)$ \cite{Oeckl2010}.
Given the state $\psi\in \cH_{\partial M}$, the subspace $L_{M} \subset L_{\partial M}$ of {\em global solutions} on $M$ and $d\nu_M$ another Gaussian measure determined by $\frac{1}{4}g_{\partial M}(\cdot,\cdot)$ \cite{Oeckl2010}, the amplitude $\rho_M$ for a Region $M$ is given by

$$\rho_M(\psi) = \int_{L_{ M}} \psi(\phi) d\nu_M(\phi)\,.$$

As anticipated in the harmonic oscillator section, in our case this is simply the pairing which provides us with a Bogoliubov transformation.

\subsection{One particle states and one point function}
\label{sec:ladder}

In this section, we briefly give the Fock space structure of $\cH_{\Sigma}$, the state space of holomorphic quantization. Creation and annihilation operators and the one-particle Hilbert space are constructed, and the one point function is evaluated.

For an element $\xi\in L_\Sigma$ define the state $p_\xi\in\cH_\Sigma$ by
\begin{equation}\label{eq:oneparticle}
p_\xi(\phi):=\frac{1}{\sqrt{2}}\{\xi,\phi\}_\Sigma .
\end{equation}
We call $p_\xi$ the normalized one-particle state corresponding to $\xi\in L_{\Sigma}$, for reasons to be clarified in the following. In \cite{Oeckl2011}, the actions of creation and annihilation operators on $\cH^h_{\Sigma}$ have been worked out as
\begin{eqnarray}
(a^\dagger_\xi\psi)(\phi) & = & p_\xi(\phi)\psi(\phi)= \frac{1}{\sqrt{2}}\{\xi,\phi\}_\Sigma\psi(\phi)\label{createdef}\\
(a_\xi\psi)(\phi) & = & \langle K_{\Sigma;\phi}, a_\xi \psi \rangle_\Sigma = \langle a^\dagger_\xi K_{\Sigma;\phi}, \psi \rangle_\Sigma\ ,\label{annihilatedef}
\end{eqnarray}
where $K_{\Sigma;\phi}$ is a coherent state around the classical solution $\phi$, for details see appendix \ref{sec-CoStat}.

Here, $\xi\in L_{\Sigma}$, and we may say that $a^\dagger_{\xi}$ and $a_{\xi}$ create and annihilate 'a particle of type $\xi$', respectively. The action of $a_{\xi}$ on coherent states is particularly simple:
\begin{equation*}
a_\xi K_{\Sigma;\phi} = \frac{1}{\sqrt{2}}\{\phi,\xi\}_\Sigma K_{\Sigma;\phi}
\label{eq:aocact}
\end{equation*}
In particular, all annihilation operators annihilate the vacuum $\psi_{0;\Sigma}(\phi) = K_{\Sigma;0} (\phi) = 1$. These operators satisfy the usual commutation relations
\begin{equation}
[a_\xi,a_\eta^\dagger]=\{\eta,\xi\}_\Sigma,\qquad [a_\xi,a_\eta]=0,\qquad [a_\xi^\dagger,a_\eta^\dagger]=0 .
\end{equation}
The creation and annihilation operators together with the vacuum state determine a Fock structure on $\cH^h_{\Sigma}$. In particular, the one particle sector is the set
\begin{equation}
\left\lbrace\frac{1}{\sqrt{2}}\{\eta,\cdot\},\eta\in L_{\Sigma}\right\rbrace = \left\lbrace p_{\eta}, \eta\in L_{\Sigma}\right\rbrace =: L_{\Sigma}^* \simeq L_{\Sigma}\,,
\end{equation}
and we were indeed justified to call $p_\xi(\phi)=\frac{1}{\sqrt{2}}\{\xi,\phi\}_\Sigma$ a one-particle-state. This is completely analogous to the Harmonic oscillator case, where the space of first excitations is given simply by $p_a(\cdot) = \{a,\cdot\}$, with $a \in \C \sim L$. 

To evaluate the one point function in the one particle sector we make use of the construction of observables in \cite{Oeckl2012}. Let us introduce the short hand, $D(\phi)=\int dx\,\sqrt{-\det g(x)}\mu(x)\phi(x)$, and the state $\eta_D$ such that $D(\cdot)=2\omega_{\partial M}(\cdot,\eta_D)$. Then the amplitude in the presence of a source term $\mu$ can be evaluated on coherent states as:
\begin{eqnarray}
 \label{eq:muamp}
 \nonumber\rho^{\mu}_M(K_{\xi})&=&\rho_M(K_\xi)\exp(iD(\hat\xi(x))e^{\frac{i}{2}D(\eta_D)-\frac{1}{2}g_{\partial M}(\eta_D,\eta_D)}\\
&=&\rho_M(K_\xi)\exp(iD(\hat\xi(x))e^{\frac{i}{2}D((1-iJ_{\partial M})\eta_D)}\,,
\end{eqnarray}
where $\hat\xi=\xi^R-i\xi^I$ with $\xi = \xi^R + J_{\partial M}\xi^I$ and $\xi^R,\xi^I \in L_{ M}$.

The one point function for the one particle sector is then simply obtained by taking the first order in $\xi$ and the first derivative with respect to $\mu$ evaluated at $\mu = 0$, which yields the simple form

\begin{equation}
\rho^{\phi(x)}(\{\xi,\cdot\}) = \hat\xi(x)\ .
\end{equation}

\subsection{Time evolution and Unitarity}
\label{sec:timeev}

Let $M$ be a region such that $\partial M = \Sigma_1 \cup \overline{\Sigma}_2$ is the disjoint union of two hypersurfaces. We can consider canonical projections $r_1: L_{ M} \rightarrow L_{\Sigma_1}$ and $r_2: L_{ M} \rightarrow L_{\Sigma_2}$\footnote{For $\phi\in L_{ M}$ a global solution on $M$, $r_1(\phi)$ is the germ of $\phi$ at $\Sigma_1$, in other words $r_1(\phi)$ is obtained by forgetting $\phi$ everywhere but in a small neighborhood around $\Sigma_1$. An equivalent statement holds for $r_2$.}. These are linear maps. If we assume these maps are invertible, we have the composition $T:= r_2\circ r_1^{-1}: L_{\Sigma_1}\rightarrow L_{\Sigma_2}$ evolves solutions at $\Sigma_1$ into solutions at $\Sigma_2$, hence it can be seen as a generalized notion of time evolution. We call $T$ unitary if

\begin{equation}\label{eq:unitaryT} J_{\Sigma_2}\circ T = T\circ J_{\Sigma_1}.\end{equation}

Note again that this defines a notion of unitarity for the classical evolution, which is of course not a necessary prerequisite for the quantum evolution to be unitary. In fact generally the classical time evolution will not be unitary in the presence of any potential or interacting terms. In \cite{Oeckl2010} it was shown that if $T$ is unitary, then there is a unitary map $U:\mathcal{H}_{\Sigma_1}\rightarrow\mathcal{H}_{\Sigma_2}$, $\Psi\mapsto\Psi\circ T^{-1}$. In particular we have $$UK_{\Sigma_1,\xi} = K_{\Sigma_2,T\xi}.$$

Obviously $T0 = 0$ since $T$ is linear. So the vacuum state on $\Sigma_1$ will evolve into the vacuum state on $\Sigma_2$, allowing us to construct a vacuum state $K_{(0,0)}$ on $\Sigma_1 \cup \Sigma_2$ which is in this sense compatible with the dynamics. Since we can extend this construction to any hypersurface $\Sigma$, there will be a global vacuum state compatible with the dynamics if $T$ is unitary. 

Note further that if $T$ is unitary, then $U$ defined in terms of it maps the one particle Hilbert space to the one particle Hilbert space, thus this can only be the case if there is no scattering or particle creation going on between hypersurfaces.

In this case $\hat\xi=\xi^R-i\xi^I$ that arises in the one point function is given explicitly as the complex solution 
\begin{equation}
 \label{eq:hatxi}                                                                                                                                                                                                                                         
 \hat\xi=r_1^{-1}\left(\frac{1}{2}(1+iJ_{\Sigma_1})\xi_1+\frac{1}{2}(1-iJ_{\Sigma_1})T^{-1}\xi_2\right)
\end{equation}
for all $\xi=(\xi_1,\xi_2)\in L_{\partial M}=L_{\Sigma_1}\times L_{\Sigma_2}$.

More generally using the unitary map $U$ the amplitude map can be expressed using the inner product in the form
\begin{equation}
 \label{eq:unitaryampl}
 \rho_{M}(\Psi_1\otimes \iota_{\Sigma_2}{\Psi}_2)=\rho_{M}(\Psi_1\otimes \overline{{\Psi}_2})=\langle U^{-1}\Psi_2, \Psi_1\rangle_{\Sigma_1} 
\end{equation}
for all $\Psi_1\in \mathcal{H}_{\Sigma_1}$ and $\Psi_2\in \mathcal{H}_{\Sigma_2}$ and $\iota_{\Sigma_2}$ given by
\begin{equation}\label{eq:defiota}
 \iota_\Sigma:\mathcal{H}_{\Sigma}\to \mathcal{H}_{\overline{\Sigma}}\,,\quad\psi\mapsto \overline{\psi}\,.
\end{equation}

\subsection{Symmetry conditions on complex structures}\label{sec:symmred}

Let us assume that we are given some symmetries of the classical spacetime region the Klein-Gordon field under consideration is defined on. Of course, we would wish to find these symmetries also represented in the quantum field theory. Imposing this as a condition can be used to reduce the ambiguity in the choice of complex structure considerably. 

Let $\{\chi_i\}$ be a collection of vector fields encoding the classical spacetime symmetries, i.e. Killing vector fields. Then we demand that the complex structure $J_\Sigma$ acting on the field configurations $\phi$ at the hypersurface $\Sigma$ commutes with the vector fields $\chi_i^\mu\partial_\mu$ as $[\chi_i^\mu\partial_\mu,J_{\sigma}]\phi(x)|_{\Sigma}=0$ for all $i$. Let $L_\Sigma^{\{i\}}\subseteq L_{\Sigma}$ be the eigenspace of $\chi_i^\mu\partial_\mu$ to the collection of eigenvalues $\{\lambda_i\}$. The complex structure $J_{\Sigma}$ must then leave $L_\Sigma^{\{i\}}$ invariant, i.e. $J_{\Sigma}L_\Sigma^{\{i\}}=L_\Sigma^{\{i\}}$ for all $i$. Hence, we can try to fix $J_{\Sigma}$ for every $i$ separately. 

Let us for example consider the case of $1+1$ dimensional Minkowski space covered by coordinates $(t,x_1)$ in which the metric takes the form $g=\text{diag}(1,-1)$. Let us assume that we are given a region $M=\Sigma\times [x_{1,l},x_{1,r}]$ in that spacetime bounded by two hyperplanes $\Sigma_l$ and $\Sigma_r$ at $x_{1,l}$ and $x_{1,r}$ respectively with $x_{1,l}<x_{1,r}$. Then the symmetries of this region are clearly time translation, time reflection and space reflection at the hyperplane at $(x_r-x_l)/2$. Hence, imposing the compatibility with the time translation invariance on $J_{\Sigma_l}$ and $J_{\Sigma_r}$ means that they must commute with the vector field $E=i\partial_t$. Imposing this condition we find that $J_{\Sigma_l}$ and $J_{\Sigma_r}$ must leave the energy eigenspaces $L^{E}_{\Sigma_l}\subset L_{\Sigma_l}$ and $L^{E}_{\Sigma_r}\subset L_{\Sigma_r}$ respectively invariant.

Thus the choice of complex structure reduces from one posed in an infinite dimensional vector space into one choice of $J_E$ per $L^E$. As $L^E$ is two dimensional, namely the modes $\exp(i (E t \pm p x))$, the complex structures allowed per energy eigenspace are the same as those of the harmonic oscillator.

We will make use of that throughout Section \ref{sec:noclick} and \ref{sec:ExperimSituation}.

\newcommand{\dd}{\text{d}}
\newcommand{\la}{\langle}
\newcommand{\ra}{\rangle}

\subsection{The Unruh-DeWitt detector}\label{sec:UdW}

We can now dicuss the coupling of an Unruh-DeWitt type detector to the GBF states we have described above. In first order perturbation theory the response amplitude of the detector at a given field configuration is provided by the one point function of the field.

Consider a harmonic oscillator of reduced mass $\sqrt{m\Omega}$, and at coupling strength $\lambda$, travelling on a world line $\gamma$ in the presence of a field transition from the vacuum to $\psi$. Th amplitude of its transition between states differing by frequency $\Omega$ can be deduced immediately from the text book treatment of Birrell and Davis \cite{BD82} as
\begin{equation}
\rho_{BD} = i \lambda\sqrt\frac{2}{m\Omega} \int \dd \tau e^{i \Omega \tau} \la \psi|\phi(\gamma(\tau))|0\ra = i \lambda\sqrt\frac{2}{m\Omega} \int \dd \tau e^{i \Omega \tau} \rho^{\phi(\gamma(\tau))}\left(|\psi\ra \otimes |0\ra\right)\ .
\end{equation}

The discussion of perturbation theory carries over to the GBF case in much the same way as in the standard treatment. A derivation from coherent states is given in appendix \ref{sec:detmod}. One key difference, as discussed above is that we specify the state on the whole boundary, that is, the one particle states are no longer necessarily of the form $|\psi\ra \otimes |0\ra$ but can be superpositions of those. Their general functional form is $\{\xi,\cdot\}$. Recall that in the presence of such a boundary state, we found the one point function to have the simple form

\begin{equation}
\rho^{\phi(x)}(\{\xi,\cdot\} ) = \hat \xi(x)\ .
\end{equation}

Inserting this into the standard result we find that the probability $P = |\rho_{BD}|^2$ from the ground state of the harmonic oscillator to the first excited state is given by
\begin{eqnarray}
 \label{eq:detprob1}
 P^{\Omega,\xi}(g\rightarrow e)&:=&\frac{2\lambda^2}{m\Omega}\int_{\tau_1}^{\tau_2} d\tau d\tau'\, e^{i\Omega(\tau-\tau')} \hat\xi(\gamma(\tau))\overline{\hat\xi(\gamma(\tau'))}\,.
\end{eqnarray}
For the corresponding deexcitation probability we have
\begin{eqnarray}
 \label{eq:detprob2}
 P^{\Omega,\xi}(e\rightarrow g)&:=&\frac{2\lambda^2}{m\Omega}\int_{\tau_1}^{\tau_2} d\tau d\tau'\, e^{-i\Omega(\tau-\tau')} \hat\xi(\gamma(\tau))\overline{\hat\xi(\gamma(\tau'))}\,.
\end{eqnarray}

Appendix \ref{sec:detmod} contains a detailed derivation of these formulas, demonstrating that we are justified to use the standard detector formula for states specified on time like hyper surfaces.

Let us return for a moment to the interpretation of these probabilities in the presence of a boundary state $\{\xi,\cdot\}$. From the derivation in \ref{sec:detmod} we know that these probabilities are normalised to the one particle sector. Thus they express the joint probability that the harmonic oscillator experiences the described transition, and that we find the field in the particular one particle configuration $\xi$, relative to the probability that we find at most one particle on the boundary, which for the given situation turns out to be one.

Recall that we always consider the case that the boundary of the region $M$ is a disjoint union $\partial M=\Sigma_1\cup\Sigma_2$. Thus we can write the vacuum state as $\psi_{M;0}=\psi_{1;0}\otimes\psi_{2;0}$. Then, we call $\psi_{1;0}$ the vacuum state on $\Sigma_1$ and $\psi_{2;0}$ the vacuum state on $\Sigma_2$, and we can consider a $\xi$ such that $a^\dagger_\xi \psi_{M;0}=\psi_{1;0}\otimes \psi_{2}$ for some $\psi_2\in\cH_{\Sigma_2}$.

With this particular configuration of the boundary field we recover the interpretation as a transition from the vacuum state on $\Sigma_1$ to a particular one particle state on $\Sigma_2$. If $\Sigma_1$ is in the past of $\Sigma_2$ the equations \eqref{eq:detprob1} and \eqref{eq:detprob2} have the interpretation of the probability that the detector transitions and that the field goes from the vacuum to the state $\psi_2$.

In order to study the probability that the detector transitions given just the fact that the field was in the vacuum state at $\Sigma_1$ we would need to sum  \eqref{eq:detprob1} or \eqref{eq:detprob2} over an orthonormal basis of one particle states of the form $\psi_{1;0} \otimes \psi_2$, as per the general form of the probability interpretation.

The central feature that makes the particular probabilities \eqref{eq:detprob1} and \eqref{eq:detprob2} so useful for our study is that as $\hat\xi$ depends very explicitly on the complex structure. For example, where applicable equation \eqref{eq:hatxi} expresses $\hat\xi$ explicitly in terms of $J$. This will make it straightforward to study the dependence of the transition probabilities on the choice of complex structure.

\section{Vacuum states on timelike hypersurfaces}\label{sec:vacuumtimelike}

In this section we will explain how one can choose a complex structure for timelike hyperplanes in Minkowski space. The main difference between the case of timelike hypersurfaces and spacelike hypersurfaces is the appearance of exponentially increasing and decreasing solutions to the Euler-Lagrange equations called evanescent waves in contrast to the usual propagating waves. For example in $1+1$-dimensional Minkowski space solutions to the Klein-Gordon equation in coordinates $(x_0,x_1)$ in which the metric takes the constant form $g=\text{diag}(1,-1)$ can be parameterized by the Fourier transform $\eta(p_0)$ as
\begin{equation}\label{eta3}
 \eta(x_0,x_1)=\int\frac{d p_0 }{(2\pi)^3 2p_1}
  \left(\eta(p_0) f(p_0,x_1)
  e^{-ip_0 x_0}+c.c.\right) ,
\end{equation}
where $p_1:= \sqrt{|p_0^2-m^2|}$ and
\begin{equation}\label{eq:cases}
 f(p_0,x_1):=\begin{cases}
  e^{i p_1 (x_1-\zeta)}=\cos(p_1 (x_1-\zeta))+i\sin(p_1 (x_1-\zeta)) &
   \text{if}\; p_0^2-m^2>0 \\
   \cosh(p_1 (x_1-\zeta))+i \sinh(p_1 (x_1-\zeta)) & \text{if}\; p_0^2-m^2 <0\,.
 \end{cases}
\end{equation}
The first type of solutions in (\ref{eq:cases}) are called propagating and the second type evanescent waves. The restriction of $\eta$ to the hypersurface $\Sigma_\zeta$ at $x_1=\zeta$ gives 
\begin{equation}
 \eta(x_0,\zeta)=\int\frac{d p_0 }{(2\pi)^3 2p_1} \left(\eta(p_0) e^{-ip_0 x_0}+c.c.\right) ,
\end{equation}
which is a parameterization of all possible field configurations on $\Sigma_\zeta$ that possess a Fourier transform. To take all of them into account we have to deal with evanescent waves.

In order to investigate the choice of complex structure on a timelike hypersurface like $\Sigma_\zeta$ we will use the Unruh-DeWitt detector model we worked out in the preceding section. We will investigate the case of propagating waves in Section \ref{sec:noclick} and evanescent waves in Section \ref{sec:ExperimSituation}. It is there that we will arrive at the main result of this article which is that the choice of complex structure for evanescent waves depends on the physical setup outside of the region of interest, and thus no general arguments can determine a unique complex structure for these modes.

\subsection{The no-click condition and the energy-momentum flux for timelike hyperplanes}
\label{sec:noclick}

In this section we show how the detector response can be used to formulate a condition that determines the complex structure for propagating waves when we claim that the complex structure may possess all the symmetries of the spacetime region. The idea is to require that all one particle states with outgoing energy flux have no probability of exciting the detector at rest.

\begin{figure}[ht]
 \begin{center}
  \includegraphics[scale=0.5]{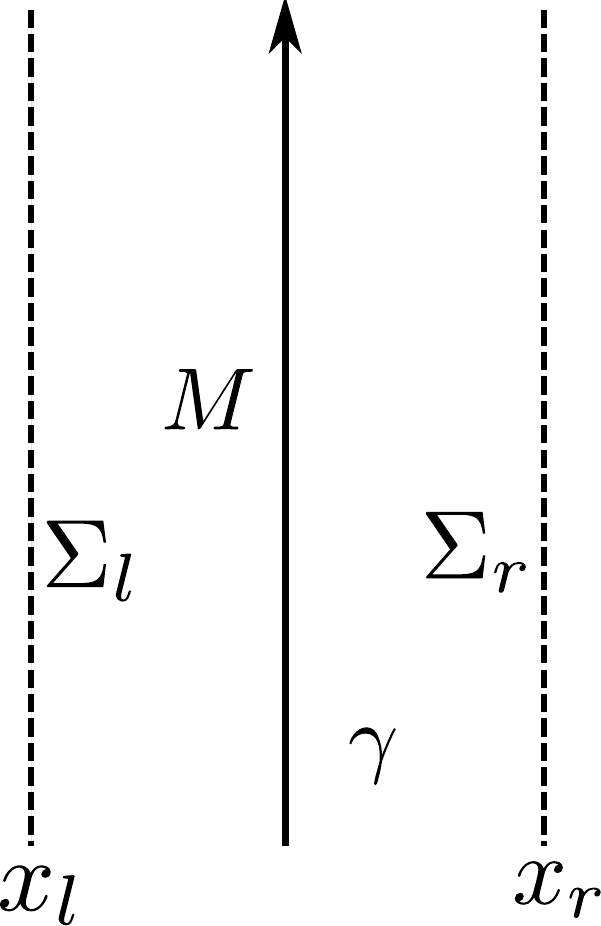}
  \caption{Schematic picture of the region $M$ containing the worldline $\gamma$ of the detector at rest. The horizontal axis represents space and the vertical axis time.  \label{fig:detectorM}}
 \end{center}
\end{figure}

We are dealing with $1+1$ dimensional Minkowski space and a region $M=\Sigma\times[x_{1,l},x_{1,r}]$ bounded by parallel timelike hyperplanes $\Sigma_l$ and $\Sigma_r$ where $\Sigma_l$ is oriented as $\Sigma$ pointing outside of the region $M$ and $\Sigma_r$ has opposite orientation. Hence, the symmetries in the global coordinates in which the time direction is tangential to the timelike hyperplanes and in which the metric takes the form $g=\text{diag}(1,-1)$ at every point of $M$ are time translation and time reversal. We use exactly this coordinate system in the following. The symplectic form is then given as
\begin{equation}
 \omega_{\partial M}(\phi,\phi')=-\frac{1}{2}\int_{-\infty}^{\infty} dx_0 (\phi\partial_{x_1}\phi'-\phi'\partial_{x_1}\phi)|_{x_{1,l}}+\frac{1}{2}\int_{-\infty}^{\infty} dx_0 (\phi\partial_{x_1}\phi'-\phi'\partial_{x_1}\phi)|_{x_{1,r}}\,.
\end{equation}
We assume the Unruh-DeWitt detector to be at rest, i.e. $\gamma(\tau)=(\tau,z)$ inside the region $M$ throughout this whole section (see figure \ref{fig:detectorM}).

The additional ingredient for determining the complex structure is the no-click condition for propagating waves which is the following requirement: We introduce the set of positive flux modes  $L^{\mathbb{C}}_{\partial M,+}\subset L^{\mathbb{C}}_{\partial M}$ consisting of solutions $\xi_{p_0}=e^{-ip_0 x_0}(ae^{ip_1x_1}+be^{-ip_1x_1})$ fulfilling
\begin{equation}\label{eq:posflux}
 -n_\mu T^{\mu0}({\mathfrak Re}(\xi_{p_0}))(x)>0
\end{equation}
at every point $x\in \partial M$ with $T^{\mu\nu}$ the energy momentum tensor of the field and $n$ the unit normal covector field to $\partial M$ pointing out of the region $M$. Now, to determine the complex structure we require that the set of states corresponding to positive flux modes should give a vanishing transition rate of the detector.\\

{\bf Comparison to the case of spacelike hypersurfaces}\\

Let us briefly consider the condition that all one particle states with outgoing energy flux have no probability of exciting the detector at rest, which we formalized above, for the case of two spacelike hypersurfaces. In that case we have a priori 4 modes per energy subspace of the complexified state space at each spacelike hypersurface, two future pointing, and two past pointing.

Out of these the complex structure picks two per surface as $+i$ eigenvalues, or, physically speaking, ``positive'' energy modes. The condition that we consider the modes that flow out of the space time then choses the past pointing modes for the past hypersurface and the future-pointing modes for the future hypersurface. On the other hand, the usual complex structure choses the future pointing modes in both cases. Thus the intersection between the two are the positive frequency modes on the future hyperplane, and we are considering vacuum to one particle transitions, which, with the detector in the ground state, cannot happen.

Any other choice of modes for the complex structure would lead to the presence of modes on the past hypersurface, and thus allow for the excitation of the detector. Thus turning the above on its head, we can use the condition that the detector should not click for outward pointing modes to uniquely fix the complex structure in the case of spacelike hypersurfaces.

Note again that we have two separate notions of positivity here, one determined by the complex structure, and the other, independently of it, by the energy flux of a certain mode.

The condition that energy should be flowing out of the region can be seen simply as a generalized version of the condition that the past hypersurface should be in the vacuum state. In particular it is a notion that survives complexification and thus can be used, in combination with the detector response, to pick the vacuum state.\\

To make the no-click condition mathematically clear we first assume that we are dealing with a unitary complex structure as defined in Section \ref{sec:timeev}. Since the complex structure respects the symmetries of the classical configuration we can deal with all the energy subspaces of $L^{\mathbb{C}}_{\Sigma}$ separately. To determine the complex structure we look at the space $L^{\mathbb{C},E}_{\Sigma}:=L^{\mathbb{C},p_0}_{\Sigma}\oplus L^{\mathbb{C},-p_0}_{\Sigma}$ with $L^{\mathbb{C},p_0}_{\Sigma}=\{\sum_{\sigma_1=\pm}a_{\sigma_1}e(p_0,\sigma_1p_1)|\,a_{\sigma_1}\in\mathbb{C}\}$ where $e(p_0,p_1)=e^{-i(p_0 x_0 - p_1 x_1)}$. Using this split we investigate the no-click condition for elements of $L^{\mathbb{C}}_{\partial M,+}\cap L^{\mathbb{C},E}_{\Sigma}$ with $L^{\mathbb{C}}_{\partial M,+}$ introduced above. We find that on $\Sigma_l$ the condition in Equation (\ref{eq:posflux}) becomes
\begin{equation}
T^{10}({\mathfrak Re}(\xi_{p_0}))(x)=  -(\partial_0 {\mathfrak Re}(\xi_{p_0}) ) (\partial_1 {\mathfrak Re}(\xi_{p_0}) )(x)>0\,,
\end{equation}
and with the opposite sign on $\Sigma_r$. We obtain that solutions in $L^{\mathbb{C}}_{\partial M,+}\cap L^{\mathbb{C},E}_{\Sigma}$ are of the form $\xi^{\mathbb{C}}_{+}=(\xi_l,\xi_r)$ with $\xi_l=c_{l,+}e(p_0,p_1)+c_{l,-}e(-p_0,-p_1)$ and $\xi_r=c_{r,+}e(p_0,-p_1)+c_{r,-}e(-p_0,p_1)$ . We define the boundary configuration to be $\xi_{+}:=\mathfrak{Re}(\xi^{\mathbb{C}}_{+})\in L_{\partial M}$. Let the coefficients $c_{i,\pm}$ be given such that $\xi_{+}$ corresponds to a normalized state $a^{\dagger}_{\xi_{+}}\psi_{M;0}$, we find that for the field being in this state the probability for the transition of a detector at rest, i.e. worldline $\gamma(\tau)=(\tau,z)$ from the ground state to the excited state following from Equation (\ref{eq:detprob1}) is given as
\begin{eqnarray}\label{eq:probposflux}
 P_\eta(g\rightarrow e)&=&\frac{2\lambda^2}{m\Omega}\int_{-\infty}^{\infty} d\tau d\tau'\, e^{i\Omega(\tau-\tau')} \hat\xi_{+}(\tau, z)\overline{\hat\xi_{+}(\tau',z)}\,\,.
\end{eqnarray}
Let us assume that the complex structure $J_{\partial M}=(J_{\Sigma_l},J_{\Sigma_r})$ is unitary (see Section \ref{sec:timeev}). Then, we have
\begin{eqnarray}\label{eq:xi_+}
  \hat \xi_{+}&=&\xi_{+}^R-i\xi_{+}^I=\frac12\left(1+iJ_{\Sigma_l}\right)\mathfrak{Re}(\xi_l)+\frac12\left(1-iJ_{\Sigma_l}\right)\mathfrak{Re}(\xi_r)\,.
\end{eqnarray}
The no-click condition is required to hold for every element of  $L^{\mathbb{C}}_{\partial M,+}$, that is for every set of coefficients $c_{l,\pm}$ and $c_{r,\pm}$. Considering the special case of $c_{l,+}=c_{l,-}$ and $c_{r,+}=c_{r,-}=0$ we obtain that 
\begin{eqnarray}
  \hat \xi_{+}&=&\xi_{+}^R-i\xi_{+}^I=c_{r,+}\left(\frac12\left(1+iJ_{\Sigma_l}\right)e(p_0,p_1)+\frac12\left(1+iJ_{\Sigma_l}\right)e(-p_0,-p_1)\right)\,.
\end{eqnarray}
which means that for the detector not to click we need
$\left(1+iJ_{\Sigma_l}\right)e(p_0,p_1)=0$ which leads to $J_{\Sigma_l}e(p_0,p_1)=ie(p_0,p_1)$ and with complex conjugation $J_{\Sigma_l}\overline{e(p_0,p_1)}=J_{\Sigma_l}e(-p_0,-p_1)=-ie(-p_0,-p_1)$. Together with the time reversal invariance which tells us that $J_{\Sigma_l}e(-p_0,p_1)=ie(-p_0,p_1)$ and $J_{\Sigma_l}e(p_0,-p_1)=-ie(p_0,-p_1)$ this already determines the complex structure to be
\begin{equation}\label{eq:Jtimelike}
 J_{\Sigma_l}=-J_{\Sigma_r}=\frac{\partial_{x_1}}{\sqrt{-\partial^2_{x_1}}}\,.
\end{equation}
Putting this back into Equation (\ref{eq:xi_+}) we find
\begin{eqnarray}
  \hat \xi_{+}=\frac12(\overline{c_{l,+}}+c_{l,-})e(-p_0,-p_1)+\frac12(\overline{c_{r,+}}+c_{r,-})e(-p_0,p_1) \,.
\end{eqnarray}
With (\ref{eq:probposflux}) we obtain that the no-click condition is fulfilled independently of the choice of the coefficients $c_{l,\pm}$ and $c_{r,\pm}$ as it was required. In particular, this includes states that can be seen as consisting of a one particle state on the right and the vacuum state on the left hypersurface and vice versa. But the above considerations were much more general as we also considered general one particle states on the whole boundary $\partial M$.\\

For evanescent wave solutions the property to be in the set of positive flux modes fulfilling condition (\ref{eq:posflux}) depends on the position of the hypersurface, i.e. for two hypersurfaces $\Sigma_\zeta$ and $\Sigma_{\zeta'}$ at $x_1=\zeta$ and $x_1=\zeta'$ respectively the set of positive flux modes is a different subset of $L^{\mathbb{C}}_{\partial M}$. In the derivation of the complex structure via the no-click condition we used explicitly that the two sets of positive flux modes for $\Sigma_l$ and $\overline{\Sigma_r}$ coincide. We conclude that the no-click condition in the form above is not applicable to evanescent waves. In the next section we will construct an explicit physical situation to show a way how to fix the complex structure for evanescent waves on timelike hypersurfaces using the Unruh-DeWitt detector and our knowledge about the complex structure on spacelike hypersurfaces.

\subsection{An explicit experimental situation with evanescent modes}\label{sec:ExperimSituation}

An example of a physical situation in which evanescent waves can be prepared is Maxwell electrodynamics in the presence of a dielectric slab. This was investigated in \cite{carniglia1971} where the authors model the experimental situation as two half spaces of different refractive index with one of them being the vacuum refractive index. In this setup solutions exist that are propagating waves in the dielectric medium and become evanescent in the vacuum. The authors of \cite{carniglia1971} find after quantization that their theoretical results coincide with semiclassical calculations and in a later publication \cite{carniglia1972} they also show that they coincide with experiments. We are going to consider a similar setup for the massive scalar field in $1+1$ dimensions. This toy model recovers the basic situation found in \cite{carniglia1971} but makes it much easier to understand the calculations which will allow us to focus on the basic insights concerning evanescent waves. We will use the Unruh-DeWitt detector as a toy model for the atoms in the electromagnetic field near the dielectric boundary. Throughout this section we will again assume the detector to be at rest, i.e. $\gamma(\tau)=(\tau,z)$.

\begin{figure}[ht]
 \begin{center}
  \includegraphics[scale=0.5]{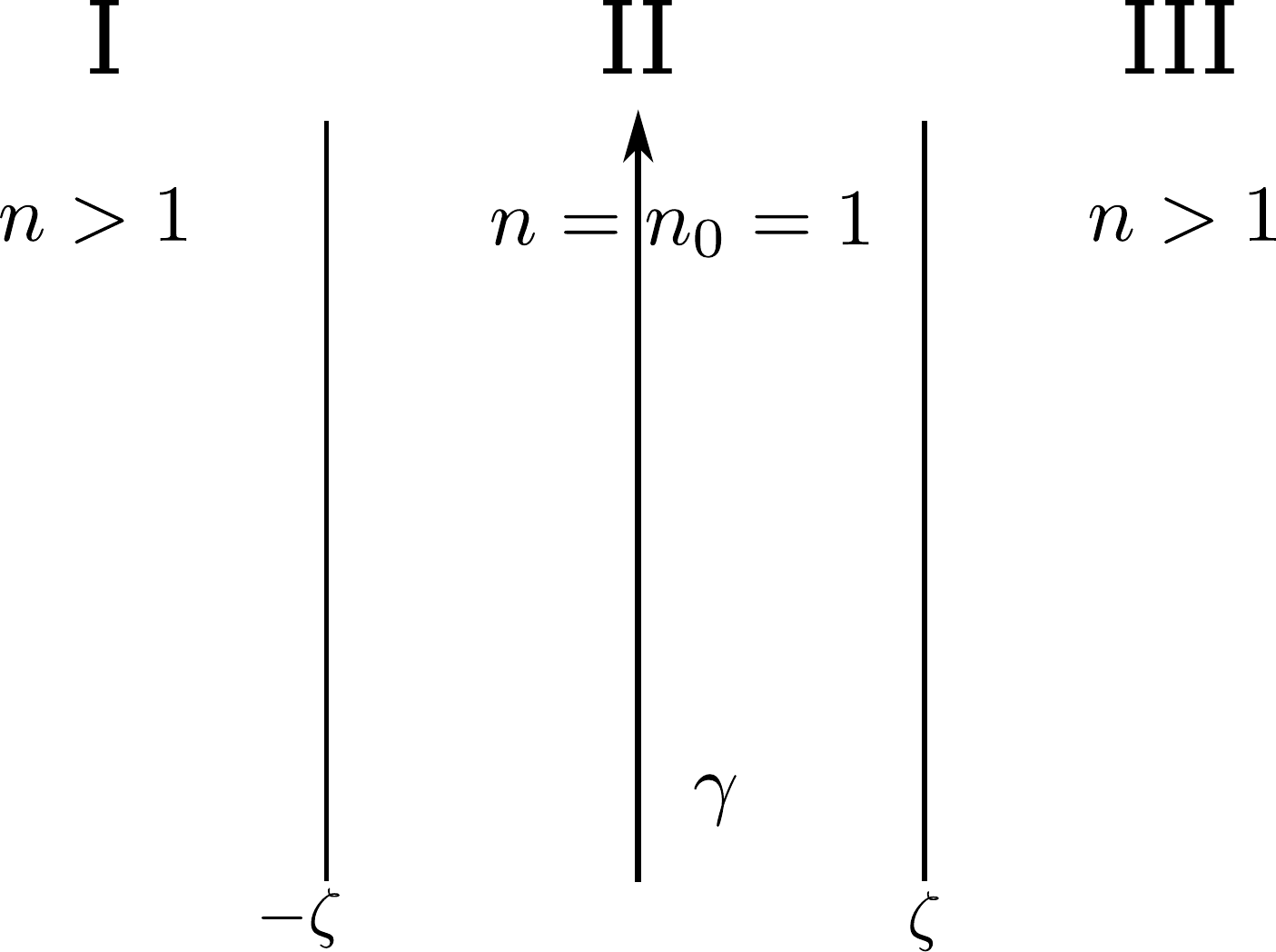}
  \caption{Schematic picture for the preparation of evanescent waves in a toy model of $1+1$-dimensional massive scalar field theory. The horizontal axis represents space and the vertical axis time.}\label{fig:refraction}
 \end{center}
\end{figure}

The setup we consider is shown in figure \ref{fig:refraction}. The $1+1$ dimensional spacetime $\Man$ covered by coordinates $(x_0,x_1)$ is separated in three regions with different metric tensor $g$. For region $I$ and $III$ we have in coordinates $$g=\left(\begin{array}{cc} \frac{1}{n} & 0 \\ 0 & -1 \end{array} \right) $$ and for region $II$ we have:

$$g=\left(\begin{array}{cc} 1 & 0 \\ 0 & -1 \end{array}\right) $$

The boundaries of the regions are at $-\zeta$ and $\zeta$. We consider the action
\begin{equation}\label{eq:KGactionslab}
 S[\phi]=\frac{1}{2}\int \xd^2x\,\sqrt{-g}\left(g^{\mu\nu}\partial_\mu\phi\partial_\nu\phi-m^2\phi^2\right)\,.
\end{equation}
for the free Klein-Gordon field. The conditions for the construction of solutions $\phi$ of the Klein-Gordon equation 
\begin{equation}\label{eq:KGeqnsslab}
 \left(\partial_\mu\sqrt{-g}\,g^{\mu\nu}\partial_\nu+\sqrt{-g}\,m^2\right)\phi(x)=0\,,
\end{equation}
at $-\zeta$ and $\zeta$ can be derived from the action as follows: Let us assume that we are given a spacetime $\Man$ and a timelike hypersurface $\Sigma$ separating $\Man$ in two regions $M_1$ and $M_2$. Let $d\sigma_{i,\mu}$ be the induced volume measure on $\Sigma_i$, and define $\Sigma_1=\Sigma$ and $\Sigma_2=\overline{\Sigma}$. By varying the action (\ref{eq:KGactionslab}) by $\delta \phi$ and partial integration we obtain
\begin{equation}
 \delta S=-\sum_i\int_{M_i} \xd^2x\,\delta\phi\left(\partial_\mu\sqrt{-g}\,g^{\mu\nu}\partial_\nu+\sqrt{-g}\,m^2\right)\phi(x)+\beta\sum_i\int d\sigma_{i,\mu}\,\left.\delta\phi \,g^{\mu\nu}\partial_\nu\phi\right|_{\Sigma_i}\,,
\end{equation}
where $\beta$ depends on the orientation of $\Sigma$. The first term leads to the Euler-Lagrange equation (\ref{eq:KGeqnsslab}) in $M_1$ and $M_2$ and the second terms leads to the condition that for $\phi$ continuous at $\Sigma$ and a jump in $g$ we must have a jump in the first derivative of $\phi$ at $\Sigma$. For the case of the dielectric slabs we obtain the boundary conditions
\begin{eqnarray}\label{eq:boundcond}
 \nonumber\lim_{\epsilon\to 0}\frac{1}{\sqrt{n}}\partial_1\phi|_{-\zeta-\epsilon}&=&\lim_{\epsilon\to 0}\partial_1\phi|_{-\zeta+\epsilon}\quad \text{and}\\
 \lim_{\epsilon\to 0}\partial_1\phi|_{\zeta-\epsilon}&=&\lim_{\epsilon\to 0}\frac{1}{\sqrt{n}}\partial_1\phi|_{\zeta+\epsilon}\,.
\end{eqnarray}
With the separation 
\begin{equation} \label{eq:wavebasis}
\phi^\pm_{p_0}(x)=\varphi^\pm_{p_0}(x_1)e^{-ip_0x_0}
\end{equation}
with $p_0>0$ we obtain for $p_0^2<m^2$  the set of solutions $\varphi^+_{p_0}(x_1)$ defined in the appendix \ref{sec:orthonormaliy} in Equation (\ref{eq:solutions}) and $\varphi^-_{p_0}(x_1)=\varphi^{+}_{p_0}(-x_1)$. These solutions are chosen such that with the symplectic form for an equal time hypersurface given as
\begin{equation*}
 \omega_{x_0}(\phi,\phi')=\frac{1}{2}\int_{-\infty}^{\infty} dx_1 \left.\sqrt{-g}\,g^{00}(\phi\partial_0\phi'-\phi'\partial_t\phi)\right|_{x_0}\,.
\end{equation*}
they fulfill the orthonormality relation with the Klein-Gordon inner product
\begin{equation}
\label{eq:orthonormalityrel}
 (\phi^s_{p_0},\phi^{s'}_{p'_0})_{KG}:=i\omega_{x_0}(\overline{\phi^{s}_{p_0}},\phi^{s'}_{p'_0})=2\pi \sqrt{n}p_0 \delta(p-p')\delta_{s,s'}\,.
\end{equation}
With Equation (\ref{eq:orthonormalityrel}) the canonical quantization prescription leads to the absorption and emission operator
\begin{equation*}
 \hat\phi(x)=\int_0^\infty\frac{dp}{2\pi  \sqrt{n}p_0}\sum_{s=\pm}\left(a_{p,s}\phi_{p_0}^s(x)+h.c.\right)
\end{equation*}
with the commutation relations $[a_{p,s}^\dagger,a_{p',s'}]=(\phi^s_{p_0},\phi^{s'}_{p'_0})_{KG}=2\pi\sqrt{n} p_0\delta(p-p')\delta_{s,s'}$. We define the momentum $p=(np_0^2-m^2)^{1/2}$ in region $I$ and $III$ and $p_1=(m^2-p_0^2)^{1/2}$ in region $II$, respectively.
The total probabilities for an Unruh-DeWitt detector with the world line $\gamma(\tau)=(\tau,z)$ to emit a field particle while passing from its ground state to an exited state and vice versa is then given as
\begin{eqnarray}\label{eq:emissionprob}
 \nonumber P^{\Omega}(g\rightarrow e)&=&\int_0^\infty\frac{dp}{2\pi  \sqrt{n}p_0}\sum_{s=\pm}\left|\langle \psi_0|a_{p,s}\frac{\lambda}{m\Omega}\int_{-\infty}^{\infty} \tau \hat\phi(\gamma(\tau)) |\psi_0\rangle\right|^2\\
\nonumber&=&\frac{\lambda^2}{m\Omega}\int_0^\infty\frac{dp}{2\pi\sqrt{n} p_0}\int_{-\infty}^\infty d\tau d\tau' e^{i(\Omega+p_0)(\tau-\tau')}\sum_{s=\pm}|\varphi_{p_0}^{s}(z)|^2 \\
 P^{\Omega}(e\rightarrow g)&=&\frac{\lambda^2}{m\Omega}\int_0^\infty\frac{dp}{2\pi\sqrt{n} p_0}\int_{-\infty}^\infty d\tau d\tau' e^{-i(\Omega-p_0)(\tau-\tau')}\sum_{s=\pm}|\varphi_{p_0}^{s}(z)|^2\,.
\end{eqnarray}
 It turns out that the probability $P^{\Omega}(g\rightarrow e)$ vanishes and with the expression in Equation (\ref{eq:solutions}) we find the transition probability
\begin{equation}\label{eq:gamma}
 \begin{array}{ll}& P^{\Omega}(e\rightarrow g):=\frac{\lambda^2 \sqrt{n}}{m\Omega p}\sum_{s=\pm}|\varphi_{\Omega}^{s}(z)|^2\\ \\
= & \frac{\lambda^2 \sqrt{n}}{m\Omega p}|N_{\Omega}|^2\frac{p}{\sqrt{n}p_1}\left(\left(\frac{\sqrt{n}p_1}{p}-\frac{p}{\sqrt{n}p_1}\right)+\right.\\ \\
 &\left.+\frac12\left(\frac{\sqrt{n}p_1}{p}+\frac{p}{\sqrt{n}p_1}\right)\left(\cosh2p_1(z + \zeta) + \cosh2p_1(z-\zeta)\right)\right)\delta(0)^2\\ \\
=&\frac{\lambda^2 }{m\Omega p_1}|N_{\Omega}|^2\left(\left(\frac{\sqrt{n}p_1}{p}-\frac{p}{\sqrt{n}p_1}\right)+\left(\frac{\sqrt{n}p_1}{p}+\frac{p}{\sqrt{n}p_1}\right)\cosh2p_1 z\,\cosh2p_1\zeta\right)\delta(0)^2\,,\end{array}
\end{equation}
that depends on the space point $z$ at which the detector rests and on the energy $\Omega$ and we have the usual divergence $\delta(0)^2$ due to the infinite duration of the interaction. If we want to know the probabilities for the absorption of a particle by the detector undergoing the same transitions as above we just have to interchange $\Omega$ by $-\Omega$ in Equation (\ref{eq:emissionprob}). Then we find that $P_{\text{absorption}}^{\Omega}(e\rightarrow g)=0$ and $P_{\text{absorption}}^{\Omega}(g\rightarrow e)=P_{\text{emission}}^{\Omega}(e\rightarrow g)$ where the latter is the transition probability in Equation (\ref{eq:emissionprob}).

\begin{figure}[ht]
 \begin{center}
  \includegraphics[scale=0.5]{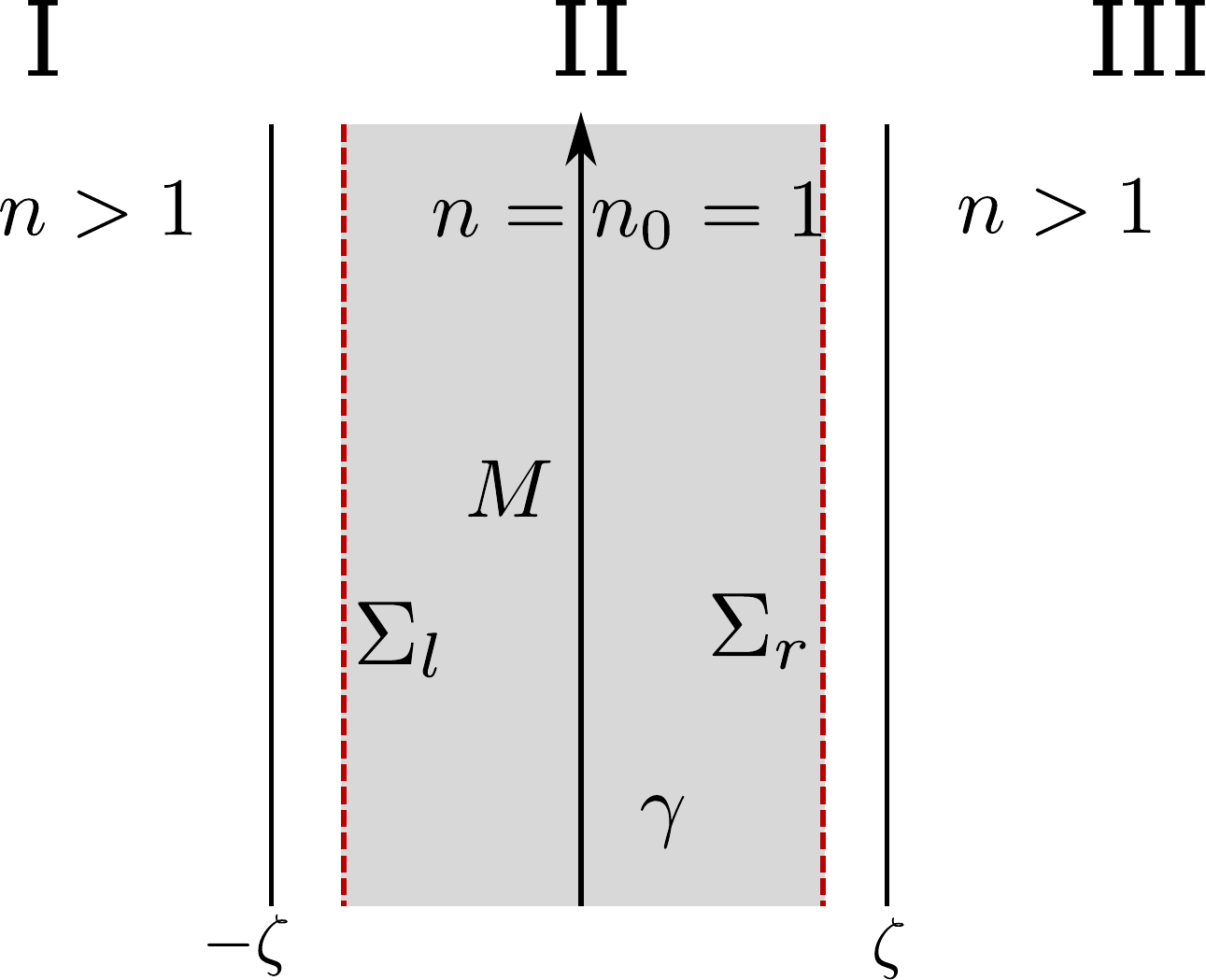}
  \caption{The region $M$ lies completely inside the region $II$ and the detector is at rest inside the region $M$. The horizontal axis represents space and the vertical axis time.}\label{fig:timelikediel}
 \end{center}
\end{figure}

\paragraph{The GBF description.} We now give a GBF description for a region in between the dielectric slabs. Take the region $M=\Sigma\times[x_{1,l},x_{1,r}]$ with timelike boundaries $\Sigma_l$ and $\Sigma_r$ with unit normal pointing outside of $M$ and $M$ lying completely in region $II$ (see figure \ref{fig:timelikediel}). 
The symplectic form is given as $\omega_{\partial M}=\omega_{\Sigma_l}+\omega_{\Sigma_r}$ where
\begin{equation}
\label{eq:1+1omegatimelike}
 \omega_{\Sigma_{l,r}}(\eta_1,\eta_2) =\frac{\epsilon_{l,r}}{2}\int dx_0\,
  \left.\left(\eta_1(x_0,x_1) \partial_{x_1} \eta_2(x_0,x_1) - \eta_2(x_0,x_1)\partial_{x_1}\eta_1(x_0,x_1)\right)\right|_{\Sigma_{l,r}}.
\end{equation}
and $\epsilon_{l}=-1$ and $\epsilon_{r}=1$.
Since the boundary of $M$ is a disjoint union $\partial M= \Sigma_l\cup \Sigma_r$ the complex structure $J_{\partial M}$ acts as a complex structure $J_{\Sigma_l}$ on $L_{\Sigma_l}$ and $J_{\Sigma_r}$ on $L_{\Sigma_r}$, respectively. 

Let us consider a detector at rest at $x_1=0$, i.e., $\gamma(\tau)=(\tau,0)$. Consider the sum $P^{\Omega,\text{tot}}(g\rightarrow e):=P^{\Omega,\text{out}}(g\rightarrow e)+P^{\Omega,\text{in}}(g\rightarrow e)$, that is, the transition probability of a detector given an undetermined one particle state on the boundary. In the canonical theory we know this to be (\ref{eq:gamma}). Note that (\ref{eq:gamma}) depends on the distance $\zeta$ from the origin to the boundary of the regions I and III and on the refraction index $n$, in other words on information about the physics outside of $M$. However, $P^{\Omega,\text{tot}}(g\rightarrow e)$ does not depend on anything but the identification of the one particle sector and the complex structure. Thus, if we accept that the restriction to the one particle sector in the GBF and the canonical case should encode the same physical situation, we already see that the only way to recover the canonical result is by selecting an appropriate complex structure which encodes this physical information. In particular the correct complex structure for a space time region cannot be given without knowledge of the physics in the spacetime into which the region in question is embedded.  In fact it is a priori unclear that it is possible to find such a complex structure.

We will now show that in the particular physical configuration we are considering in this section it is indeed possible to recover the canonical result in Equation (\ref{eq:gamma}) by an appropriate choice of complex structure for the boundary of the region $M$. As before, the complex structures $J_l$ and $J_r$ can be specified by giving the corresponding eigenfunctions $e_{p_0;l}e^{ip_0 x_0}\in L^{\C}_{\Sigma_l}$ and $e_{p_0;\Sigma_R}e^{ip_0 x_0}\in L^{\C}_{\Sigma_r}$ such that $J_l e_{p_0;l}e^{ip_0 x_0}=ie_{p_0;l}e^{ip_0 x_0}$ and $J_r e_{p_0;r}e^{ip_0 x_0}=ie_{p_0;r}e^{ip_0 x_0}$.

We will show that the complex structure given by setting $e_{p_0;l}(x_1):=c_{a;l}(p_0)\cosh(p_1 x_1)+c_{b;l}(p_0)\sinh(p_1 x_1)$ with the coefficients
\begin{eqnarray}\label{eq:jldef}
  \nonumber c_{a;l}(p_0)&=&\cosh(p_1\zeta)+i\frac{p}{\sqrt{n}p_1}\sinh(p_1\zeta)\\
   c_{b;l}(p_0)&=&\sinh(p_1\zeta)+i\frac{p}{\sqrt{n}p_1}\cosh(p_1\zeta)\,,
\end{eqnarray}
and by setting $e_{p_0;r}(x_1):=c_{a;r}(p_0)\cosh(p_1 x_1)+c_{b;r}(p_0)\sinh(p_1 x_1)$ with
\begin{eqnarray}\label{eq:jrdef}
  \nonumber c_{a;r}(p_0)&=&\cosh(p_1\zeta)-i\frac{p}{\sqrt{n}p_1}\sinh(p_1\zeta)\\
   c_{b;_r}(p_0)&=&-\sinh(p_1\zeta)+i\frac{p}{\sqrt{n}p_1}\cosh(p_1\zeta)\,,
\end{eqnarray}
indeed reproduces the canonical result (\ref{eq:gamma}) from Equation (\ref{eq:probgeneral}). In particular, this complex structure is non-unitary.\\

We obtain the probability for the transition of the detector from the ground state to the excited state by summing the expression in Equation (\ref{eq:detprob1}) over a set of mutually orthogonal, normalized states spanning the set of possible states of the scalar field, i.e., by summing over the set representing our ignorance of the final state of the system. 

As in the canonical case, we assume that there are no field particles going in the region $M$. In the canonical case, this was done by fixing the initial state on the initial hypersurface to be the vacuum state. This is more complicated in the case of a timelike hypersurface. However, we can use the transition probability of the detector itself to decide which states correspond to ingoing and outgoing particles by comparing the result to the canonical case: the transition probability from the ground state to the excited state of the detector was zero in the canonical case. From Equation (\ref{eq:detprob1}) with $\gamma(\tau)=(\tau,0)$, we obtain that the one particle states $a^\dagger_{\xi}\psi_{M;0}$ in $\cH_{\partial M}$ that are given such that $\hat\xi\propto e^{ip_0x_0}$ with $p_0> 0$ and $\hat\xi$ derived from $\xi$ as defined in Equation (\ref{eq:hatxi}) lead to a vanishing transition probability. Hence, they are the outgoing states. Conversely, the ingoing one particle states are linear combinations of the states $a^\dagger_{\xi}\psi_{M;0}$ given such that $\hat\xi\propto e^{ip_0x_0}$ with $p_0< 0$. 

We obtain that for two one particle states we have
\begin{equation}
 \label{eq:stateprod}
 \langle a^{\dagger}_\xi \psi_{M;0}, a^{\dagger}_{\xi'} \psi_{M;0}\rangle =\{\xi',\xi\}_{\partial M}=g_{\partial M}\left(\hat\xi',\overline{\hat\xi}\right)\,.
\end{equation}
where the symmetric, bilinear map $g_{\partial M}$ is extended to $L^{\mathbb{C}}_{\partial M}$ as $g_{\partial M}(\phi+i\phi',\psi):=g_{\partial M}(\phi,\psi)+ig_{\partial M}(\phi',\psi)$ for real $\phi,\phi',\psi$. Hence, two one particle states corresponding to $\xi$ and $\xi'$ are orthogonal if and only if the corresponding $\hat\xi$ and $\hat\xi'$ are orthogonal with respect to $g_{\partial M}$. 

The complex structure $J_{\partial M}$ defined in equations (\ref{eq:jldef}) and (\ref{eq:jrdef}) leaves the energy eigenspaces
$$L_{M}^{p_0,\mathbb{C}}:=\{c_+ e^+_{p_0;l}+c_- \overline{e^+_{-p_0;l}}|\,c_+,c_-\in\mathbb{C}\}$$
invariant. Hence, $J_{\partial M}$ maps ingoing states to ingoing states and outgoing states to outgoing states, and we find from the definition of $g_{\partial M}$ from $\omega_{\partial M}$ and $J_{\partial M}$ that ingoing and outgoing states are orthogonal. 

Let us now consider just the ingoing states to show that we recover the canonical result in Equation (\ref{eq:gamma}) for the transition of the detector from its excited state to its ground state: since the complex structure defined by equations (\ref{eq:jldef}) and (\ref{eq:jrdef}) leaves the energy eigenspaces $L_{M}^{p_0,\mathbb{C}}$ invariant, we see from Equation (\ref{eq:stateprod}) that two ingoing states $a^\dagger_{\xi}\psi_{0;M}$ and $a^\dagger_{\xi'}\psi_{0;M}$ are orthogonal if $\hat\xi\propto e^{ip_0x_0}$ and $\hat\xi'\propto e^{ip'_0x_0}$ with $p_0\neq p'_0$. Hence, we obtain an orthonormal set of ingoing one particle states by specifying a set of orthonormal one particle states $a^\dagger_{\xi_1}\psi_{M;0}$ and $a^\dagger_{\xi_2}\psi_{M;0}$ corresponding to solutions $\xi_1,\xi_2$ in $L_{M}^{p_0,\mathbb{C}}$ for every $p_0$. 

From Equation (\ref{eq:stateprod}) we know that the orthonormality of the one particle states $a^\dagger_{\xi_1}\psi_{0;M}$ and $a^\dagger_{\xi_2}\psi_{0;M}$ is equivalent to the orthonormality of $\hat\xi_1$ and $\hat\xi_2$ with respect to the sesquilinear map $(\cdot,\cdot):=g_{\partial M}(\cdot ,\overline{\cdot })$ which is anti-linear in the second entry. We start with the two linearly independent eigenfunctions of $J_{\Sigma_l}$ defining $\xi^{+}_{p_0}:=e^+_{p_0;l}$ and $\xi^{-}_{p_0}:=\overline{\xi^{+}_{-p_0}}$. We define the normalization constant $c_{p_0}^{+}$ such that $(\xi^{+}_{p_0},\xi^{+}_{p'_0})=4\pi (c_{p_0}^{+})^{-1}\delta(p_0-p'_0)$. An orthonormal basis in the one particle sector is then given by the set $a^\dagger_{\xi_1}\psi_{0;M}$ and $a^\dagger_{\xi_2}\psi_{0;M}$, where 
\begin{eqnarray}
\label{eq:schmidt}
 \hat\xi_1=\tilde{\xi}^{+}_{p_0}&:=&\sqrt{c_{p_0}^{+}}\xi^{+}_{p_0}\\
 \hat\xi_2=\tilde{\xi}^{-}_{p_0}&:=&\sqrt{c_{p_0}^{-}}\left(\xi^{-}_{p_0}-\frac{(\xi^{-}_{p_0},\xi^{+}_{p_0})}{(\xi^{+}_{p_0},\xi^{+}_{p_0})}\xi^{+}_{p_0}\right)\,,
\end{eqnarray}
with $\sqrt{c_{p_0}^{-}}$ such that $(\tilde{\xi}^{-}_{p_0},\tilde{\xi}^{-}_{p_0})=4\pi \delta(p_0-p'_0)$.
From Equation (\ref{eq:detprob2}) we find the corresponding response of the detector as
\begin{equation}
 \label{eq:probgeneral}
 P^{\Omega,p_0}(e\rightarrow g)=\frac{\lambda^2}{m\Omega}\int_{-\infty}^{\infty} d\tau d\tau'\, e^{-i(\Omega+p_0)(\tau-\tau')} \left(|\tilde{\xi}^{+}_{p_0}(z)|^2+|\tilde{\xi}^{-}_{p_0}(z)|^2\right)\,.
\end{equation}
Now, we insert the eigenfunctions of the complex structure defined in equations (\ref{eq:jldef}) and (\ref{eq:jrdef}) and recover the expression in Equation (\ref{eq:gamma}) as it is shown in Appendix \ref{sec:canres}.

While this is not the only possible complex structure here, it is very instructive to see what happens if we use the equations of motion to unitarily propagate the hypersurface $\Sigma_l$ and $\Sigma_r$ and the corresponding complex structures to the regions $I$ and $III$ respectively by considering only the global solution defined in (\ref{eq:solutions}) which are in one-to-one correspondence with field configurations on $\Sigma_l$ and $\Sigma_r$, respectively\footnote{That has been already done for the case of empty $3+1$-dimensional Minkowski space and a region bounded by a timelike hyper-cylinder in \cite{Colosi2008,Colosi2008a}.}.

Employing the boundary conditions in Equation (\ref{eq:boundcond}), we obtain for the global solutions
\begin{equation}\label{eq:unitpropl}
 e^+_{p_0;l}(x_0,x_1)=\left\{\begin{array}{lcr}  \Big(e_{p_0;l}(-\zeta)\cos p(x_1+\zeta)+&&\\+\frac{\sqrt{n}}{p}\left.\left(\frac{d}{dx_1}e_{p_0;l}(x_1)\right)\right|_{x_1=-\zeta}\sin p(x_1+\zeta)\Big)e^{ip_0 x_0}
& \text{for}  & x_1<-\zeta \\  e_{p_0;l}(x_1)e^{ip_0 x_0} & \text{for} & -\zeta\le x_1\le \zeta \,,\end{array}\right.
\end{equation}
 and 
\begin{equation}\label{eq:unitpropr}
 e^+_{p_0;r}(x_0,x_1)=\left\{\begin{array}{lcr}   e_{p_0;r}(x_1)e^{ip_0 x_0} & \text{for} & -\zeta\le x_1\le \zeta \\ \Big(e_{p_0;r}(\zeta)\cos p(x_1-\zeta)+&&\\+\frac{\sqrt{n}}{p} \left.\left(\frac{d}{dx_1}e_{p_0;r}(x_1)\right)\right|_{x_1=\zeta}\sin p(x_1-\zeta)\Big)e^{ip_0 x_0} &\text{for}& \zeta< x_1  \,.\end{array}\right.
\end{equation}
Note also that, as in the canonical theory, the extensions to the regions $I$ and $III$ of $e^+_{p_0;l}$ and $e^+_{p_0;r}$ respectively are only continuous, but not analytic at $x_1 = \pm\zeta$ since the spacetime metric $g$ has a discontinuity at $-\zeta$ respectively $\zeta$. Thus in particular an arbitrarily small region in spacetime will no longer automatically contain the complete information on all spacetime without recurse to the equations of motion. Mathematically it is this fact that allows us to study the dependence of the complex structure on the physics outside of the spacetime region.

Through Equation (\ref{eq:unitpropl}), \eqref{eq:jldef}, (\ref{eq:unitpropr}) and \eqref{eq:jrdef} respectively we see that the corresponding unitarily propagated complex structure in region $I$ and $III$ coincides with the complex structure given in Equation (\ref{eq:Jtimelike}), 

\begin{equation}
J_{\Sigma_l}=-J_{\Sigma_r}=\frac{\partial_{x_1}}{\sqrt{-\partial^2_{x_1}}}\,.
\end{equation}

This is just the complex structure that we found to be the natural one for propagating waves in Section \ref{sec:noclick}. Thus the complex structure we gave above for evanescent waves is directly related to a natural complex structure for the propagating waves that they turn into asymptotically. This further confirms our interpretation of the complex structure \eqref{eq:jldef}, \eqref{eq:jrdef} as the correct one for these evanescent modes, and thus that the complex structure, and therefore the vacuum state, on timelike boundaries can depend on the physics outside of the space time region they bound.

\section{Conclusions}

In this article we introduced an Unruh-DeWitt detector model in the framework of the general boundary formulation (GBF) of quantum field theory. The Unruh-DeWitt detector is frequently used in quantum optics and quantum field theory as a toy model for the interaction of fields with matter systems like atoms. In particular, the Unruh-DeWitt detector is used to model situations of accelerated observers moving in flat spacetimes or general inertial or non-inertial observers moving in curved spacetimes. Usually, this is done to gain a better understanding of effects that predict the particle creation in the vacuum like the Unruh effect or the Hawking effect. The introduction of an Unruh-DeWitt detector in the GBF allows for the investigation of such questions employing the advantages of the GBF framework: there is no restriction of the spacetime region that can be considered. Especially, regions with timelike hypersurfaces can be considered, e.g. in the case of a region with timelike boundary completely outside of the horizon of an eternal black hole could be considered. 

Beside this, the Unruh-DeWitt detector we introduced could be used to investigate the general boundary quantum field theory in deSitter space and anti-deSitter space for which unique vacuum states and S-matrices were introduced in \cite{Col10} and \cite{Colosi2011a}, respectively.

In this paper, we used the Unruh-DeWitt detector model to show that the transition rates of a detector when coupled to a scalar field are an appropriate tool to determine the complex structure for propagating wave modes of the scalar field. In order to study this physical situation we had to re-express the condition that we have a transition from the vacuum to the one particle sector in GBF language. We saw that considering energy fluxes was a viable method in our case. However, we only studied these for the highly symmetric and simple case of a region in Minkowski space bounded by parallel timelike hyperplanes. An interesting open problem then is to understand the fully quantized energy momentum tensor and its detailed dependence on the complex structure. This would allow us to handle a much richer class of physical phenomena.

We considered evanescent modes of a scalar field in a particular spacetime configuration representing a toy model for the physical situation of the electromagnetic field in the vacuum between two dielectric slabs. We found that even if the spacetime region under consideration is isometric to a region in Minkowski space the complex structure depends on the physics outside the spacetime region, where the metric might differ from the Minkowskian. In this sense we cannot describe the physics in a spacetime region purely locally without considering the embedding of the region into a spacetime. 

This result should be contrasted with the classic result of Reeh and Schlieder \cite{Reeh:Schlieder:1961,0138.45301,2008arXiv0802.1854S} in algebraic quantum field theory. They showed that the vacuum state of Minkowski space is a highly non-local structure. To compare to the language of algebraic quantum field theory we could describe the physical situation considered above as such: Consider two different static spacetimes, which coincide in a certain region. If we model the vacuum state of the second spacetime in the first one by applying operators localized outside of the region where the spacetimes coincide, we will generically change the vacuum state inside the coinciding region, too. The restrictions of the vacuum states to the coinciding spacetime region do not coincide, which is exactly what we found. As the response of the detector to evanescent modes is exponentially decaying away from the dielectric boundary, we have in some sense that, as in AQFT, the non-locality is exponentially small.

Note further that though it may seem at first sight that this example violates the generally covariant locality principle \cite{Brunetti:2001dx}, it is only the vacuum state that exhibits this non-locality, not the observables. 

\section*{Acknowledgements}

DR thanks Robert Oeckl, Daniele Colosi and Max Dohse for very helpful discussions, especially during a visit at the UNAM Campus Morelia. This research stay was funded by CONACYT grant 49093. The authors thank Robert Oeckl and Benjamin Bahr for reading the manuscript and many helpful remarks. The authors thank Wojciech Kami\'nski for very helpful discussions and for pointing out how to find the right boundary conditions for a relativistic field at a metric discontinuity. The work of RB and DR has been supported by the International Max Planck Research School for Geometric Analysis, Gravitation and String Theory. 

\appendix

\section{Appendix}

\subsection{Coherent States}\label{sec-CoStat}

In this appendix we briefly review the notion of coherent states for the holomorphic Klein-Gordon field. Fixing any element $\xi \in L_{\Sigma}$, the corresponding coherent state $K_{\Sigma;\xi}\in\mathcal{H}_{\Sigma}$ is the holomorphic function
\begin{equation}
 K_{\Sigma;\xi}(\phi) := \exp\left(\frac{1}{2}\{\xi,\phi\}_{\Sigma}\right)\qquad \forall\phi\in L_{\Sigma}\,.
 \label{eq:coherent_state}
\end{equation}
Note that if we reverse the orientation of $\Sigma$, $\Sigma \rightarrow \overline{\Sigma}$, then the symplectic form is mapped to $\omega_{\overline{\Sigma}} = - \omega_{\Sigma}$ and $g_{\overline\Sigma} = g_{\Sigma}$ and therefore $J_{\overline{\Sigma}}=-J_\Sigma$ to preserve the positivity. Then we obtain that
\begin{equation}
\overline{\{\xi,\phi\}}_{\Sigma} = g_\Sigma(\xi,\phi) - i \omega_\Sigma(\xi,\phi) = g_{\overline\Sigma}(\xi,\phi) + i \omega_{\overline\Sigma}(\xi,\phi) = \{\xi,\phi\}_{\overline\Sigma}\,,
\end{equation}
and, thus, $K_{\Sigma;\xi}$ is mapped to $K_{\overline{\Sigma};\xi}=\overline{K_{\Sigma;\xi}}$. This justifies the identification of the Hilbert spaces $\mathcal{H}_{\Sigma}$ and $\mathcal{H}_{\overline{\Sigma}}$ by the map $\iota$ in \eqref{eq:defiota}.

Remark that if $\Sigma$ decomposes into two components, $\Sigma = \Sigma_1 \cup \bar{\Sigma}_2$, then for any $\xi \in L_{\Sigma_1}$ and $\xi' \in L_{\Sigma_2}$, we have
\begin{equation} 
 \label{eq:coherentdecomp}
 K_{\Sigma;(\xi,\xi')} = K_{\Sigma_1;\xi}\otimes K_{\overline{\Sigma_2};\xi'}\in \mathcal{H}_{\Sigma}\,.
\end{equation}

With respect to the inner product defined above, a coherent state has the reproducing kernel property $\langle K_{\Sigma;\xi},\psi\rangle_{\Sigma} = \psi(\xi)\;\forall\psi\in\mathcal{H}_{\Sigma}$. We define $\tau^R,\tau^I \in L_{ M}$ such that $\tau = \tau^R + J_{\partial M}\tau^I$ which is always uniquely possible since $L_{\partial M} = L_{ M} \oplus J_{\partial M}L_{M}$, see Proposition 4.2 in \cite{Oeckl2010}. Then, we obtain for the amplitude
\begin{equation}
\label{eq:freeamplitude}
 \rho_M(K_{\Sigma;\tau}) = \exp\left(\frac{1}{4}g_{\partial M}(\tau^R,\tau^R) - \frac{1}{4}g_{\partial M}(\tau^I,\tau^I)-\frac{i}{2}g_{\partial M}(\tau^R,\tau^I)\right)\,.
\end{equation}

There is a distinguished coherent state on the hypersurface $\Sigma$,
\begin{equation}
\psi_{\Sigma;0}:=K_{\Sigma;0} = 1
\end{equation}
which is the coherent state associated with the vector $0\in L_{\Sigma}$. This state has a natural interpretation as the vacuum state on $\Sigma$. In particular, it satisfies all vacuum axioms posed in \cite{Oeckl2010}. Note that while the vacuum state is uniquely determined here, its actual functional form depends on the complex structure. The problem of defining a vacuum state has thus been moved into determining the complex structure, where it can be stated as the problem of choosing the positive frequency modes. For more on the equivalence between the two perspectives on the problem of the vacuum see \cite{Oeckl2011b}.

\subsection{Unruh-DeWitt detector derivation in the general boundary formulation.}
\label{sec:detector}\label{sec:detmod}\label{sec:PertGBF}

In this section we will develop the standard perturbation theory for the GBF in terms of coherent states. and use this to explicitly write the first order coupling of a particle detector.

As detector we will use the model of the harmonic oscillator we developed in Section \ref{sec:harmosc}. In order to study its coupling to the free scalar field, we will introduce interaction terms as Weyl observables into the free amplitudes. We then develop the perturbation theory for these Interaction terms to first order. This all happens in strict analogy to the standard quantum field theoretic treatment, though our formalism and context are conceptually more general.

We assume that the detector moves on a worldline $\gamma_0:M=[\tau_i,\tau_f]\subset \R\to \Man$ where $\Man$ represents spacetime. The boundary of $M$ is then $\partial M=\{\tau_i\}\cup\overline{\{\tau_f\}}$. To recover the standard quantization for the harmonic oscillator we use as in Section \ref{sec:harmosc} the complex structure $J_{\tau}$ with
\begin{equation}
 J_{\tau}=-\frac{1}{\Omega}\frac{d}{d\tau}\,.
\end{equation}
We model the interaction of the harmonic oscillator $q$ with the field $\phi$ by considering the classical action
\begin{equation*}
 S_{\mathcal{V}_{UD}}[\phi,q]=S_{0;1}[\phi]+S_{0;2}[q]+\mathcal{V}_{UD}(\phi,q)\,,
\end{equation*}
where $S_{0;1}[\phi]$ and $S_{0;2}[q]$ are the free actions for the field and the harmonic oscillator, respectively, and the interaction term is given as \cite{lin2007} 
\begin{equation}
\label{eq:UD}
 \mathcal{V}_{UD}(\phi,q)=\lambda\int d^2x\,\int d\tau\, \delta^{(2)}(x-\gamma(\tau))\phi(x)q(\tau)=\lambda \int d\tau \phi(\gamma(\tau))q(\tau)\,.
\end{equation}
Let us express the amplitude for the interacting theory using the path integral:
\begin{equation}\label{eq:amplitudeVUD}
 \rho^{\mathcal{V}_{UD}}(\psi)=\int \xD\phi\xD q\, \psi(\phi,q) e^{iS_{\mathcal{V}_{UD}}[\phi,q]}\,.
\end{equation}
Let $\cH_1$ be the Hilbert space for the free quantum field $\phi$ and $\cH_2$ be the Hilbert space for the harmonic oscillator $q$. We define the actions
\begin{eqnarray}
 S_{\mu_1}[\phi]=S_{0;1}[\phi]+\int d^2x \mu_1(x)\phi(x)\\
 S_{\mu_2}[q]=S_{0;2}[q]+\int d\tau \mu_2(\tau)q(\tau)\,,
\end{eqnarray}
where $\mu_1$ and $\mu_2$ are test functions defined on the domain of $\phi$ and $q$, respectively. Then, the expression in Equation (\ref{eq:amplitudeVUD}) can be rewritten assuming that the state $\psi$ splits as $\psi=\psi_1\otimes \psi_2\in \cH_1\otimes \cH_2$ as
\begin{eqnarray}\label{eq:amplitudegenerated}
 \rho^{\mathcal{V}_{UD}}(\psi)&=& e^{i\mathcal{V}_{UD}(-i\frac{\delta }{\delta \mu_1},-i\frac{\delta}{\delta \mu_2})}\int \xD\phi\xD q \,\psi(\phi,q) e^{i(S_{\mu_1}[\phi]+S_{\mu_2}[q])}\nonumber
\\
  &=&e^{i\mathcal{V}_{UD}(-i\frac{\delta }{\delta \mu_1},-i\frac{\delta}{\delta \mu_2})}\rho^{\mu_1}(\psi_1)\rho^{\mu_2}(\psi_2)\,,
\end{eqnarray}
where $\rho^{\mu_1}$ and $\rho^{\mu_2}$ are the corresponding amplitudes sometimes called generating functionals. The trick to use generating functionals is well known from the standard formulation of quantum field theory. We generalize this here to the GBF by defining the amplitude for the interacting theory in the GBF as \cite{Colosi2008}
\begin{equation}
 \rho^{\mathcal{V}_{UD}}_M(\psi):=e^{i\mathcal{V}_{UD}(-i\frac{\delta }{\delta \mu_1},-i\frac{\delta}{\delta \mu_2})}\rho^{\mu_1}_M(\psi_1)\rho_{\gamma_0}^{\mu_2}(\psi_2)\,.
\end{equation}
In particular, to calculate the amplitude $\rho_{M}^{\mu_1}$ we can consider the additional term $\int d^2x \mu_1(x)\phi(x)$ in the corresponding action $S_{\mu_1}[\phi]$ as a linear observable $D_1$ giving rise to a Weyl observable: $W_1(\phi)=\exp(iD_1(\phi))=\exp(i\int d^2x \mu_1(x)\phi(x))$. This is again motivated from the path integral formulation.
Hence, we can use Equation (\ref{eq:muamp}) for the observable amplitude of a Weyl observable. 

By considering the harmonic oscillator as a $0$-dimensional quantum field theory, i.e. a quantum field theory at a point, we can use a similar construction with the Weyl observable $W_2(q)=\exp(iD_2(q))=\exp(i\int d\tau \mu_2(\tau)q(\tau))$. Then, we obtain for the boundary state $K_q:=K_{(q_1,q_2)}=K_{q_1}\otimes \overline{K_{q_2}}$ with 
\begin{equation*}\label{eq:boundarystate}
 q_i(\tau)=\frac{1}{\sqrt{4m\Omega}}\left(a_ie^{-i\Omega\tau}+\overline{a_i}e^{i\Omega\tau}\right)\,.
\end{equation*}
which is the parameterization introduced for the harmonic oscillator in Section \ref{sec:compcompl}, using the unitarity of $J_{\partial \gamma_0}$:
\begin{eqnarray}
\nonumber \rho^{\mu_2}_{\gamma_0}(K_{q_1}\otimes \overline{K_{q_2}})&=&\exp\left(\frac{1}{2}\overline{a_1}a_2+\frac{i}{\sqrt{4m\Omega}}\int d\tau\,\mu_2(\tau)\left(a_1e^{-i\Omega\tau}+\overline{a_2}e^{i\Omega\tau}\right)\right.\\
&&\left.+\frac{i}{2}\int d\tau d\tau'\,\mu_2(\tau)\left(1-iJ_{\partial \gamma_0}\right)\eta_{D,2}(\tau')\right)\,,
\end{eqnarray}
with $\eta_{D,2}$ the unique element of $J_{\partial \gamma_0}L_{\gamma_0}$ such that $D_2(q)=2\omega_{\partial\gamma_0}(q,\eta_{D,2})$ for all $q\in L_{\gamma_0}$.

Now, we cannot calculate the full expression in Equation (\ref{eq:amplitudeVUD}), instead, we define the term of perturbative order $n$ for the coherent states $K_{\xi}$ for the field and $K_{q}=K_{q_1}\otimes \overline{K_{q_2}}$ for the detector, respectively, as
\begin{equation}
 \label{eq:smatrix}
 \rho^{\mathcal{V}_{UD}}_{M;n}(K_{\xi}\otimes K_{q}):=\left.\frac{1}{n!}\left(\mathcal{V}_{UD}\left(-i\frac{\delta}{\delta\mu_1},-i\frac{\delta}{\delta\mu_2}\right)\right)^n\rho^{\mu_1}_M(K_{\xi_1})\rho^{\mu_2}_{\gamma_0}(K_{q})\right|_{\mu_1,\mu_2=0}\,.
\end{equation}
The amplitude in Equation (\ref{eq:smatrix}) is the analog of the scattering matrix of $n$-th order in standard QFT. 

In particular, we are interested in the first order which is given as
\begin{eqnarray*}
 &&\rho^{\mathcal{V}_{UD}}_{M;1}(K_{\xi}\otimes K_{q})\\
&=&\left.\mathcal{V}_{UD}\left(-i\frac{\delta}{\delta\mu_1},-i\frac{\delta}{\delta\mu_2}\right)\rho^{\mu_1}_M(K_{\xi})\rho^{\mu_2}_{\gamma_0}(K_{q})\right|_{\mu_1,\mu_2=0}\,.
\end{eqnarray*}
The amplitude $\rho^{\mu_\phi}_M(K_{\xi})$ is given by Equation (\ref{eq:muamp}) as
\begin{equation}
 \rho^{\mu_1}_M(K_{\xi})=\rho_M(K_\xi)e^{iD(\hat\xi)+\frac{i}{2}D((1-iJ_{\partial M})\eta_D)}
\end{equation}
with $\eta_D \in L_{\partial M}$ such that $D(\phi)=2\omega_{\partial M}(\phi,\eta_D)$. Then, with the fact that the only term of first order in the test function $\mu_1$ is $D(\hat\xi)$, we obtain with the definition of the potential in Equation (\ref{eq:UD}):
\begin{eqnarray*}
 &&\rho^{\mathcal{V}_{UD}}_{M;1}(K_{\xi}\otimes K_{q})\\
 &=& -\lambda\rho_{\gamma_0}(K_q) \rho_M(K_{\xi})\int d\tau\,\hat{q}(\tau)\hat\xi(\gamma(\tau)) \\
&=&-\lambda \rho_M(K_{\xi})e^{\frac{1}{2}\overline{a_1}a_2}\frac{1}{\sqrt{4m\Omega}}\int d\tau\,\left(a_1e^{-i\Omega\tau}+\overline{a_2}e^{i\Omega\tau}\right)\hat\xi(\gamma(\tau))\,.
\end{eqnarray*}
Expressing one particle states as first derivatives of coherent states, we derive
\begin{eqnarray}\label{eq:generalOscillatorTransition}
 &&\rho^{\mathcal{V}_{UD}}_{M;1}(a^\dagger_\xi \psi_0\otimes a^\dagger_q \psi_0)=\left.2\frac{d}{d\alpha}\frac{d}{d\beta}\rho^{\mathcal{V}_{UD}}_{M;1}(K_{\alpha\xi}\otimes K_{\beta q})\right|_{\alpha,\beta=0}\nonumber\\
&=&\frac{\lambda}{\sqrt{m\Omega}}\int_{\tau_1}^{\tau_2} d\tau\,\left(a_1e^{-i\Omega\tau}+\overline{a_2}e^{i\Omega\tau}\right)\hat\xi(\gamma(\tau))\,.
\end{eqnarray}

To interpret this amplitude, recall from Equation (\ref{eq:GBFProbability}) that in the GBF the probability to find the state in some subspace $\mathcal J$ with basis $\xi_a$, given that it was in some subspace $\mathcal I$ with basis $\xi'_b$ can be written as
\begin{equation}\label{eq:transprobGeneral}
 P(\mathcal J|\mathcal I)=\frac{\sum_{\xi_a}|\sum_n\rho^{\mathcal{V}_{UD}}_{M;n}(\xi_a)|^2}{\sum_{\xi'_b}|\sum_n\rho^{\mathcal{V}_{UD}}_{M;n}(\xi'_b)|^2}\,.
\end{equation}
Let $\psi\in \mathcal I$ be a normalized one particle state. Then the probability for the system to be found in the state $\psi$ is given by taking $\mathcal I$ the subspace spanned by $\psi$. To first non-vanishing order in the perturbation theory this is given by
\begin{equation}
 P(\psi|\mathcal I)=\frac{|\rho^{\mathcal{V}_{UD}}_{M;1}(\psi)|^2}{\sum_{\xi'_b}|\rho^{\mathcal{V}_{UD}}_{M;0}(\xi'_b)|^2}\,,
\end{equation}
where $\rho^{\mathcal{V}_{UD}}_{M;0}$ is just the free amplitude and we used that a one particle state is a linear functional and  $\rho^{\mathcal{V}_{UD}}_{M;0}$ is given as a Gaussian integral over an even domain and hence always vanishes for a one particle state which is an odd function of the field. Let $\mathcal I$ be the sector of the boundary Hilbert space spanned by the vacuum and the one particle states, that is, $\{c\psi_{\partial M;0}|\,c\in\C\}\oplus L_{\partial M}$ understood as subspaces of $\cH_{\partial M}$, i.e. we consider only states that contain at most one particle, we find that
\begin{equation}
\label{eq:transprob}
 P(\psi|\mathcal I)=|\rho^{\mathcal{V}_{UD}}_{M;1}(\psi)|^2\,.
\end{equation}
since the free amplitude for a one particle state vanishes and the amplitude for the vacuum state is always one by normalization. From Equation (\ref{eq:transprob}) we know that the absolute square of $\rho^{\mathcal{V}_{UD}}_{M;1}(a^\dagger_\xi \psi_0\otimes a^\dagger_q \psi_0)$ can be interpreted as a transition probability within the one particle sector.

We are interested in the transition of the detector from the ground state $\psi_0$ at $\tau_1$ to a normalized, excited state $\psi_e(q)$ at $\tau_2$ and vice versa at boundary field state  $a^\dagger_\xi \psi_0$, for some fixed field configuration $\xi$. Using the definition of one particle states in Section \ref{sec:ladder}, the normalized, excited state of the harmonic oscillator at a component of the boundary is easily seen to be 
\begin{equation}
 \psi_e(q)=p_{q_e}(q)=\frac{1}{\sqrt{2}}\{q_e,q\}\,,
\end{equation}
with
\begin{equation}
 q_e=\frac{1}{\sqrt{2m\Omega}}\left(e^{i\Omega\tau}+e^{-i\Omega\tau}\right)\,.
\end{equation}

In order to study the transition from the ground state at $\tau_1$ to an excited state at $\tau_2$ we set $a_1=0$ and $a_2=1$ in \eqref{eq:generalOscillatorTransition} to obtain the boundary state $\psi_0(q_1) \otimes \psi_e(q_2)$. We then find that the transition described by this boundary together with the one particle field state $a^\dagger_\xi \psi_{M;0}\in \cH_{M}$ has the probability
\begin{eqnarray}
 P^{\Omega,\xi}(g\rightarrow e)&:=&\frac{2\lambda^2}{m\Omega}\int_{\tau_1}^{\tau_2} d\tau d\tau'\, e^{i\Omega(\tau-\tau')} \hat\xi(\gamma(\tau))\overline{\hat\xi(\gamma(\tau'))}\,.
\end{eqnarray}
With $a_1=1$ and $a_2=0$ we obtain for the transition from the excited to the ground state
\begin{eqnarray}
  P^{\Omega,\xi}(e\rightarrow g)&:=&\frac{2\lambda^2}{m\Omega}\int_{\tau_1}^{\tau_2} d\tau d\tau'\, e^{-i\Omega(\tau-\tau')} \hat\xi(\gamma(\tau))\overline{\hat\xi(\gamma(\tau'))}\,.
\end{eqnarray}

Which are precisely the transition probabilities \eqref{eq:detprob1} and \eqref{eq:detprob2}.

\subsection{Orthonormality}
\label{sec:orthonormaliy}

The solutions to the Klein-Gordon equation in the setup of Section \ref{sec:ExperimSituation} are:
\begin{equation}
\label{eq:solutions}
 \varphi^+_{p_0}(x_1)=\left\{\begin{array}{lcr} N_{p_0} e^{-ipx_1}& \text{for} & x_1<-\zeta \\ N_{p_0}e^{ip\zeta}\left(\cosh{p_1 (x_1+\zeta)}+B_{p_0}\sinh{p_1 (x_1+\zeta)}\right) & \text{for} & -\zeta\le x_1\le \zeta \\ e^{-ipx_1}+C_{p_0} e^{ipx_1} &\text{for}& \zeta< x_1  \end{array}\right.\,,
\end{equation}
with
\begin{eqnarray}
\label{eq:normalization}
 \nonumber N_{p_0}&=&e^{-2ip\zeta}\left(\cosh2p_1\zeta+\frac{1}{2}\left(B_{p_0}+B_{p_0}^{-1}\right)\sinh2p_1\zeta\right)^{-1}\\
 \nonumber B_{p_0}&=&-i\frac{p}{\sqrt{n}p_1}\\
 C_{p_0}&=&N_{p_0}\frac{1}{2}\left(B_{p_0}-B_{p_0}^{-1}\right)\sinh2p_1\zeta\,,
\end{eqnarray}
and $p=(np_0^2-m^2)^{1/2}$ and $p_1=(m^2-p_0^2)^{1/2}$. In this section we will show the orthogonality relation we use in section \ref{sec:ExperimSituation}.

First of all, we show that solutions of different energy are orthogonal in any case. This we see from the time translation invariance of the symplectic structure:
\begin{equation*}
 0=\partial_0\omega_{x_0}(\overline{\phi^s_{p_0}},\phi^r_{p'_0})=i(p_0-p'_0)\omega_{x_0}(\overline{\phi^s_{p_0}},\phi^r_{p'_0})=(p_0-p'_0)(\phi^s_{p_0},\phi^r_{p'_0})
\end{equation*}
which means for $p_0\neq p'_0$ that $(\phi^s_{p_0},\phi^r_{p'_0})$ must vanish.
Now, Equation (\ref{eq:orthonormalityrel}) can be derived using the continuity equation $\partial_\mu j^\mu(\phi,\phi')=0$. We define $j^\mu(\phi,\phi'):=\sqrt{g}\,g^{\mu\nu}(\phi\partial_\nu\phi'-\phi'\partial_\nu\phi)$, and then 
\begin{equation}
 j^0(\overline{\phi^s_{p_0}},\phi^r_{p'_0})=i\frac{1}{p_0-p'_0}\partial_1j^1(\overline{\phi^s_{p_0}},\phi^r_{p'_0})
\end{equation}
which leads to
\begin{eqnarray*}
 \omega_{x_0}(\overline{\phi^s_{p_0}},\phi^r_{p'_0})&=&\frac{i}{2}\lim_{\sigma\rightarrow\infty} \left.\frac{1}{p_0-p'_0}j^1(\overline{\phi^s_{p_0}},\phi^r_{p'_0})(x) \right|_{x_1=-\sigma}^{x_1 = \sigma}\\
&=&\frac{i}{2}\lim_{\sigma\rightarrow\infty} \left.\frac{n}{p-p'}\frac{p_0+p'_0}{p+p'}j^1(\overline{\phi^s_{p_0}},\phi^r_{p'_0})(x) \right|_{x_1 = -\sigma}^{x_1=\sigma}\,.
\end{eqnarray*}
Then, we plug in the functions given in (\ref{eq:solutions}) and neglect all terms that come with the factor $p+p'$ since they vanish in the limit $\sigma\rightarrow 0$. Using the identities $\overline{C_{p_0}}N_{p_0}+C_{p_0}\overline{N_{p_0}}=0$ and $|C_{p_0}|^2+|N_{p_0}|^2=1$ and the representation of the delta function
\begin{equation*}
 \lim_{\sigma\rightarrow \infty}\frac{\sin (p-p')\sigma}{p-p'}=\pi\delta(p-p')
\end{equation*}
we arrive at
\begin{eqnarray*}
 \omega_{x_0}(\overline{\phi^+_{p_0}},\phi^-_{p'_0})&=&0\\
 \omega_{x_0}(\overline{\phi^+_{p_0}},\phi^+_{p'_0})&=&i\sqrt{n}2\pi p_0\delta(p-p')
\end{eqnarray*}
which leads to Equation (\ref{eq:orthonormalityrel}).

\subsection{Recovering the canonical result with the non-unitary complex structure}
\label{sec:canres}

First, we find that we can express the eigenfunctions of the complex structures of $J_{\Sigma_l}$ and $J_{\Sigma_r}$ in terms of each other by the relation
\begin{equation}
 \binom{e^+_{p_0,l}}{\overline{e^+_{-p_0,l}}}=\left(\begin{array}{cc} A(p_0) & B(p_0) \\ \overline{B(p_0)} & \overline{A(p_0)} \end{array}\right)\binom{e^+_{p_0,r}}{\overline{e^+_{-p_0,r}}}=:T\binom{e^+_{p_0,r}}{\overline{e^+_{-p_0,r}}}\,,
\end{equation}
where 
\begin{eqnarray*}
 A(p_0)&=&-\frac{i}{2}\left(\frac{\sqrt{n}p_1}{p}-\frac{p}{\sqrt{n}p_1}\right)\sinh 2p_1\zeta+ \cosh 2p_1 \zeta\\
 B(p_0)&=&\frac{i}{2}\left(\frac{\sqrt{n}p_1}{p}+\frac{p}{\sqrt{n}p_1}\right)\sinh 2p_1\zeta
\end{eqnarray*}
and we find that
\begin{equation}
 T^{-1}=\left(\begin{array}{cc} \overline{A(p_0)} & -B(p_0) \\ -\overline{B(p_0)} & A(p_0) \end{array}\right)\,.
\end{equation}
From that we calculate for $\xi^+_{p_0}= e^+_{p_0,\Sigma_r}$ and $\xi^-_{p_0}=\overline{e^+_{-p_0,\Sigma_r}}$ 
\begin{eqnarray*}
 (\xi^{+}_{p_0},\xi^{+}_{p'_0})=g_{\Sigma_l}(\xi^{+}_{p_0},\overline{\xi^{+}_{p'_0}})+g_{\Sigma_r}(\xi^{+}_{p_0},\overline{\xi^{+}_{p'_0}})\\
 =(|A(p'_0)|^2+|B(p'_0)|^2+1)g_{\Sigma_r}(\xi^{+}_{p_0},\overline{\xi^{+}_{p'_0}})=(|A(p_0)|^2+|B(p_0)|^2+1)2\pi p\delta(p_0-p'_0)\,
\end{eqnarray*}
and the same way
\begin{eqnarray*}
 (\xi^{-}_{p_0},\xi^{+}_{p'_0})&=&g_{\Sigma_l}(\xi^{-}_{p_0},\overline{\xi^{+}_{p'_0}})+g_{\Sigma_r}(\xi^{-}_{p_0},\overline{\xi^{+}_{p'_0}})\\
&=&-2A(p_0)\overline{B(p_0)}\,g_{\Sigma_r}\left(\overline{e^+_{-p_0,r}},e^+_{-p_0,r}\right)=-2A(p_0)\overline{B(p_0)}2\pi p\delta(p_0-p'_0)\,
\end{eqnarray*}
which leads us to
\begin{eqnarray*}
 (c_{p_0}^{+})^{-1}&=&\frac{p}{\sqrt{n}p_1}(|A(p_0)|^2+|B(p_0)|^2+1)\\
&=&\frac{p}{4\sqrt{n}p_1}\left(6-\frac{np_1^2}{p^2}-\frac{p^2}{np_1^2} +\left(\frac{\sqrt{n}p_1}{p}+\frac{p}{\sqrt{n}p_1}\right)^2\cosh 4p_1\zeta\right)  \,.
\end{eqnarray*}
Furthermore, we define
\begin{eqnarray*}
 g_{p_0}&:=&-2\frac{p}{\sqrt{n}p_1}A(p_0)\overline{B(p_0)}\\
&=&\frac{p}{2\sqrt{n}p_1}\left(\frac{\sqrt{n}p_1}{p}+\frac{p}{\sqrt{n}p_1}\right)\sinh 2p_1\zeta\left(\left(\frac{\sqrt{n}p_1}{p}-\frac{p}{\sqrt{n}p_1}\right)\sinh 2p_1\zeta+2i\cosh 2p_1\zeta\right)
\end{eqnarray*}
and obtain 
\begin{equation}
 (c_{p_0}^{-})^{-1}=(c_{p_0}^{+})^{-1}\left(1-(c_{p_0}^{+})^2|g_{p_0}|^2\right)=2\frac{p}{\sqrt{n}p_1}\,.
\end{equation}
With the definition in Equation (\ref{eq:schmidt}) we have
\begin{eqnarray*}
 \tilde{\xi}^{+}_{p_0}&:=&\sqrt{c_{p_0}^{+}}\xi^{+}_{p_0}\\
 \tilde{\xi}^{-}_{p_0}&:=&\sqrt{c_{p_0}^{-}}\left(\xi^{-}_{p_0}-c_{p_0}^{+}g_{p_0}\xi^{+}_{p_0}\right)\\
 &=&\sqrt{\frac{\sqrt{n}p_1}{2p}}\left(\cosh 2p_1\zeta+\frac{i}{2}\left( \frac{np_1^2}{p^2}-\frac{p^2}{np_1^2}\right)\sinh 2p_1 \zeta\right)^{-1}\times\\
&&\times\left(\cosh p_1(x+\zeta)-i\frac{p}{\sqrt{n}p_1}\sinh p_1(x+\zeta)\right)
\end{eqnarray*}
and arrive at
\begin{eqnarray*}
 |\tilde{\xi}^{+}_{p_0}|^2+|\tilde{\xi}^{-}_{p_0}|^2=\frac{1}{2}|N_{p_0}|^2\left(\left(\frac{\sqrt{n}p_1}{p}-\frac{p}{\sqrt{n}p_1}\right)+\left(\frac{\sqrt{n}p_1}{p}+\frac{p}{\sqrt{n}p_1}\right)\cosh2p_1 x_1\,\cosh2p_1\zeta\right)
\end{eqnarray*}
with 
\begin{equation}
 |N_{p_0}|^{-2}=\frac{1}{8}\left(6-\frac{np_1^2}{p^2}-\frac{p^2}{np_1^2} +\left(\frac{\sqrt{n}p_1}{p}+\frac{p}{\sqrt{n}p_1}\right)^2\cosh 4p_1\zeta\right)
\end{equation}
which shows that we recover the canonical result.

%
%

\end{document}